\documentclass[prd,preprint,showpacs,preprintnumbers,nofootinbib,eqsecnum,superscriptaddress]{revtex4}

 \usepackage[dvips,final]{graphicx}
  \usepackage{amssymb}
   \usepackage{amsmath}
    \usepackage{amsfonts}
     \usepackage{epsfig}
      \usepackage{bm}

\usepackage{mathpazo}

\usepackage[section]{placeins}

\usepackage{multirow}
\usepackage{ctable}
\usepackage{booktabs}
\usepackage{array}
\usepackage{tabularx}
\usepackage{xcolor}
\usepackage{pstricks}

\begin{document}

\title{Prompt inclusive production of $J/\psi$, $\psi'$ and $\chi_{c}$ mesons at the LHC \\
in forward directions within the NRQCD $k_t$-factorization approach -
search for the onset of gluon saturation}

\author{Anna Cisek}
\email{acisek@ur.edu.pl}
\affiliation{Faculty of Mathematics and Natural Sciences,
University of Rzesz\'ow, ul. Pigonia 1, PL-35-310 Rzesz\'ow, Poland}

\author{Antoni Szczurek}
\email{antoni.szczurek@ifj.edu.pl} 
\affiliation{Institute of Nuclear Physics, Polish Academy of Sciences, 
ul. Radzikowskiego 152, PL-31-342 Krak{\'o}w, Poland}
\affiliation{Faculty of Mathematics and Natural Sciences,
University of Rzesz\'ow, ul. Pigonia 1, PL-35-310 Rzesz\'ow, Poland}

\begin{abstract}
We discuss prompt production of $J/\psi$ mesons in proton-proton
collisions at the LHC within NRQCD $k_t$-factorization approach
using Kimber-Martin-Ryskin (KMR) unintegrated gluon distributions (UGDF).
We include both direct color-singlet production ($g g \to J/\psi g$) 
as well as a feed-down from $\chi_c \to J/\psi \gamma$ and 
$\psi' \to J/\psi X$ decays.
The production of the decaying mesons ($\chi_c$ or $\psi'$) is also 
calculated within NRQCD $k_t$-factorization approach.
The corresponding matrix elements for $g g \to J/\psi$ g, $g g \to \psi'g$
and $g g \to \chi_c$ include parameters of the nonrelativistic space 
wave functions of the quarkonia at $r = 0$,
which are taken from potential models in the literature.
We get the ratio of the corresponding cross section ratio for 
$\chi_c(2)$-to-$\chi_c(1)$ at midrapidities much closer to experimental data than
obtained in a recent analysis.
Differential distributions in rapidity and transverse momentum
of $J/\psi$ and $\psi'$ are calculated and compared with
experimental data of the ALICE and LHCb collaborations.
We discuss a possible onset of gluon saturation effects in the production of
$J/\psi$ and $\chi_{c}$ mesons at forward/backward rapidities.
We show that it is neccessary to modify the standard KMR UGDF
to describe ALICE and LHCb data. A mixed UGDF scenario was proposed.
Then we can describe the experimental data for $J/\psi$
production within model uncertainties with color-singlet component only.
Therefore our theoretical results leave only a relatively small room for 
the color-octet contributions.
\end{abstract}

\maketitle

\section{Introduction}
There is a long-standing lack of convergence in understaning
production of $J/\psi$ quarkonia in proton-proton or proton-antiproton
collisions. Some authors believe that the corresponding
cross sections are dominated by the so-called color-octet contribution.
On the other hand some other authors expect that the color-singlet 
contribution dominates. The color-octet contribution cannot 
be calculated from first principle and is rather fitted 
to the experimental data.
The fits lead to different size of the color-octet contribution,
depending on the details of calculations of the color-singlet
contribution(s). In many cases successful fits were obtained
but, in our opinion, there is no clear understanding of the problem.
Different fits from the literature give different magnitude
of the color-octet contributions classified according to 
quantum numbers of the $c \bar c$ system.

In the present paper we wish to calculate the color-singlet contributions 
in the NRQCD $k_t$-factorization approach and see how much 
room is left for the more difficult color-octet contribution.

It is known that a sizeable part of the $J/\psi$ production comes
from radiative decays of $\chi_c$ mesons. Therefore in the following
we have to include also this contribution very carefully trying
to confront with experimental data for $\chi_{c}$ production whenever possible.

In a very recent $k_t$-factorization analysis of $\chi_c$ production
\cite{BLZ2015b} the authors found very different values of 
the nonrelativistic wave function at the origin for 
$\chi_c(1)$ and $\chi_c(2)$:
\begin{equation}
|R_{\chi_c{(1)}}'(0)|^2 \approx 5 |R_{\chi_c{(2)}}'(0)|^2 
\end{equation}
from a fit based on $k_t$-factorization approach to LHC data.
This large modification would put in doubts either NRQCD approach 
and/or validity of the leading-order $k_t$-factorization.
In the standard potential model one obtains the same radial wave
function for different $\chi_c$ species \cite{EQ1993}.
Here we wish to discuss also this element of the whole construction.
In the following we shall use Kniehl-Vasin-Saleev matrix elements
which are given explicitly in Ref.\cite{KVS2006}.

Finally $\psi'$ quarkonium also has a sizeable branching fraction
into $J/\psi X$ \cite{PDG}. Fortunately this contribution is much
smaller than the direct one as will be discussed in this paper.
It was considered recently in almost identical approach in
\cite{BLZ2015a}.

In the present approach we concentrate rather on small transverse
momenta of $J/\psi$ or $\psi'$ relevant for ALICE and LHCb data
\cite{ALICE_2011_7TeV,ALICE_2012_2760,ALICE_2014_7TeV,LHCb_2011_7TeV,LHCb_2015_13TeV}. 
We expect that color-singlet contributions may dominate in 
this region of the phase space.

\section{Some theoretical aspects}

In the following we shall consider only color-singlet mechanisms and look how much
room is left for color-octet production.

\subsection{Main contributions}

\begin{figure}
\includegraphics[width=6.5cm]{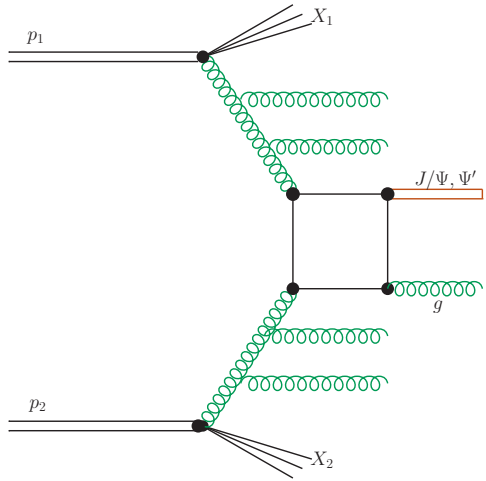}
\caption{The leading-order diagram for direct $J/\psi$ ($\psi'$) 
meson production in the $k_t$-factorization approach.}
\label{fig:gg_Jpsig}
\end{figure}

The main color-singlet mechanism of $J/\psi$ meson production is illustrated
in Fig.\ref{fig:gg_Jpsig}. In this case $J/\psi$ is produced
in association with an extra ``hard'' gluon due to C-parity conservation.

We calculate the dominant color-singlet $g g \to J/\psi g$
contribution taking into account transverse momenta of initial gluons.
In the $k_t$-factorization approach the differential cross section can 
be written as:
\begin{eqnarray}
\frac{d \sigma(p p \to J/\psi g X)}{d y_{J/\psi} d y_g d^2 p_{J/\psi,t} d^2 p_{g,t}}
&& = 
\frac{1}{16 \pi^2 {\hat s}^2} \int \frac{d^2 q_{1t}}{\pi} \frac{d^2 q_{2t}}{\pi} 
\overline{|{\cal M}_{g^{*} g^{*} \rightarrow J/\psi g}^{off-shell}|^2} 
\nonumber \\
&& \times \;\; 
\delta^2 \left( \vec{q}_{1t} + \vec{q}_{2t} - \vec{p}_{H,t} - \vec{p}_{g,t} \right)
{\cal F}_g(x_1,q_{1t}^2,\mu^2) {\cal F}_g(x_2,q_{2t}^2,\mu^2)  ;
\label{kt_fact_gg_jpsig}
\end{eqnarray}
where ${\cal F}_g$ are unintegrated (or transverse-momentum-dependent) 
gluon distributions.
The matrix elements were calculated as was explained e.g. in \cite{Baranov2002,BS2008}.
The corresponding matrix element squared for the $g g \to J/\psi g$ is
\begin{equation}
|{\cal M}_{gg \to J/\psi g}|^2 \propto \alpha_s^3 |R(0)|^2 \; .
\label{matrix_element} 
\end{equation}
Running coupling constants are used in the present calculation. 
Different combinations of renormalization scales were tried. 
We decided to use:
\begin{equation}
\alpha_s^3 \to \alpha_s(\mu_1^2) \alpha_s(\mu_2^2) \alpha_s(\mu_3^2) \; ,
\end{equation}
where $\mu_1^2 = q_{1t}^2$,
      $\mu_2^2 = q_{2t}^2$ and
      $\mu_3^2 = m_t^2$ (prescription 1)
      or
      $\mu_1^2 = max(q_{1t}^2,m_t^2)$,
      $\mu_2^2 = max(q_{2t}^2,m_t^2)$ and
      $\mu_3^2 = m_t^2$ (presciption 2),
where here $m_t$ is the $J/\psi$ transverse mass.
The factorization scale in the calculation was taken as
$\mu_F^2 = (m_t^2 + p_{t,g}^2)/2$.

\begin{figure}
\begin{center}
\includegraphics[width=6.5cm]{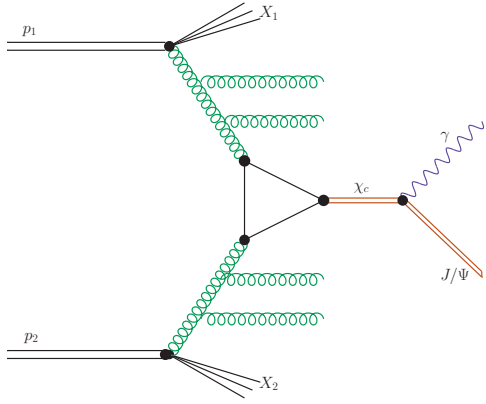}
\end{center}
\caption{The leading-order diagram for $\chi_c$ meson production
in the $k_t$-factorization approach.}
\label{fig:gg_chic}
\end{figure}

Similarly we do calculation for P-wave $\chi_c$ meson production.
Here the lowest-order subprocess $g g \to \chi_c$ is allowed by
positive $C$-parity of $\chi_c$ mesons.
In the $k_t$-factorization approach the leading-order cross section 
for the $\chi_c$ meson production can be written somewhat formally as:
\begin{eqnarray}
\sigma_{pp \to \chi_c} = \int \frac{dx_1}{x_1} \frac{dx_2}{x_2}
\frac{d^2 q_{1t}}{\pi} \frac{d^2 q_{2t}}{\pi} 
&&\delta \left( (q_1 + q_2)^2 - M_{\chi_c}^2 \right) 
\sigma_{gg \to H}(x_1,x_2,q_{1},q_{2}) \nonumber \\
&&\times \; {\cal F}_g(x_1,q_{1t}^2,\mu_F^2) {\cal F}_g(x_2,q_{2t}^2,\mu_F^2)
\; ,
\label{chic_kt_factorization}
\end{eqnarray}
where ${\cal F}_g$ are unintegrated (or transverse-momentum-dependent) 
gluon distributions and $\sigma_{g g \to \chi_c}$ is 
$g g \to \chi_c$ (off-shell) cross section.
The situation is illustrated diagramatically in Fig.\ref{fig:gg_chic}.

The matrix element squared for the $g g \to \chi_c$ subprocess is
\begin{equation}
|{\cal M}_{gg \to \chi_c}|^2 \propto \alpha_s^2 |R'(0)|^2 \; .
\label{matrix_element} 
\end{equation}

After some manipulation: 
\begin{eqnarray}
\sigma_{pp \to \chi_c} = \int d y d^2 p_t d^2 q_t \frac{1}{s x_1 x_2}
\frac{1}{m_{t,\chi_c}^2}
\overline{|{\cal M}_{g^*g^* \to \chi_c}|^2} 
{\cal F}_g(x_1,q_{1t}^2,\mu_F^2) {\cal F}_g(x_2,q_{2t}^2,\mu_F^2) / 4
\; ,
\label{useful_formula}
\end{eqnarray}
that can be also used to calculate rapidity and transverse
momentum distribution of the $\chi_c$ mesons.

In the last equation:
$\vec{p}_t = \vec{q}_{1t} + \vec{q}_{2t}$ is transverse momentum 
of the $\chi_c$ meson
and $\vec{q}_t = \vec{q}_{1t} - \vec{q}_{2t}$ is auxiliary variable 
which is used for the integration of the cross section. Furthermore:
$m_{t,{\chi_c}}$ is the so-called $\chi_c$ transverse mass and
$x_1 = \frac{m_{t,\chi_c}}{\sqrt{s}} \exp( y)$,  
$x_2 = \frac{m_{t,\chi_c}}{\sqrt{s}} \exp(-y)$.
The factor $\frac{1}{4}$ is the Jacobian of transformation from
$(\vec{q}_{1t}, \vec{q}_{2t})$ to $(\vec{p}_t, \vec{q}_{t})$ variables.

As for the $J/\psi$ production running coupling contants are used. Different
combination of scales were tried. The best choices are:
\begin{equation}
\alpha_s^2 \to \alpha_s(\mu_1^2) \alpha_s(\mu_2^2) \; ,
\end{equation}
where $\mu_1^2 = q_{1t}^2$ and
      $\mu_2^2 = q_{2t}^2$ (prescripton 1)
      or
      $\mu_1^2 = max(q_{1t}^2,m_t^2)$ and
      $\mu_2^2 = max(q_{2t}^2,m_t^2)$ (prescription2).
Above $m_t$ is transverse mass of the $\chi_c$ meson.

The factorization scale(s) for the $\chi_c$ meson production are 
fixed traditionally as $\mu_F^2 = m_t^2$.

The $J/\psi$ mesons are produced then by the $\chi_c \to J/\psi \gamma$
decays which are dominated by E1 transitions \cite{E1_1,E1_2}. 
This channel cannot be easily eliminated experimentally as 
the produced photons are usually rather soft. Due to the same reasons
$\chi_c$ mesons can be measured at large transverse momenta or 
very forward/backward directions.
 
\subsection{Unintegrated gluon distributions}

In the present analysis the Kimber-Martin-Ryskin KMR UGDFs \cite{KMR2001} 
are used, which are generated from conventional collinear MMTH2014LO PDFs \cite{HMMT2015}.
In actual calculations of distributions we interpolate them on 
a three-dimensional grid
in $log_{10}(x)$, $log_{10}(k_t^2)$ and $log_{10}(\mu^2)$ prepared before
the calculation of the production cross section or differential
distributions.

The KMR UGDF was succesfully used e.g. for production of charm and charmed mesons
\cite{MS1,LMS} as well as for production of two pairs of $c \bar{c}$ \cite{MS2,HMS}.

In standard approach the KMR UGDFs are calculated for larger values
of gluon transverse momenta and are usually frozen at small
gluon transverse momenta. The value at which the freezing is applied
is independent of all other variables, longitudinal momentum fraction
in particular.
The UGDFs used in calculations neglect possible effects of saturation.
For small initial gluon momenta $k_{1t}^2 < Q_s^2$ or $k_{2t}^2 < Q_s^2$
and for forward/backward production some effects of gluon saturation may
be expected. The saturation scale as is often parametrized as:

\begin{equation}
Q_s^2(x) = Q_0^2 \left( x_0/x \right)^{\lambda} \; .
\label{saturation_momentum}
\end{equation}
One could correct the original KMR distributions by assuming
saturation of UGDFs for $k_{it}^2 < Q_s^2$
\begin{equation}
{\cal F}_{A}(x,k_t^2,\mu^2) = const \;  for \;  k_t^2 < Q_s^2 \; .
\label{KMR_A}
\end{equation}
%
We shall call this model of UGDF ``saturation A'' for brevity.
For comparison we shall consider also faster damping of the small-$k_t$ 
region by multiplying the ${\cal F}_A$ by an extra damping factor:
%
\begin{equation}
{\cal F}_{B}(x,k_t^2,\mu^2) = \left( k_t^2 / Q_s^2 \right)
{\cal F}_{A}(x,k_t^2,\mu^2) \; .
\label{KMR_B}
\end{equation}
We shall call this model of UGDF ``saturation B'' for brevity.
Another, called ``mixed UGDF'', scenario is discussed in the text.
Some consequences of the small-$k_t$ corrections will be discussed
in the following section.

\section{Results}

\subsection{$\psi'$ production}

\begin{figure}[t]
\includegraphics[width=5.5cm]{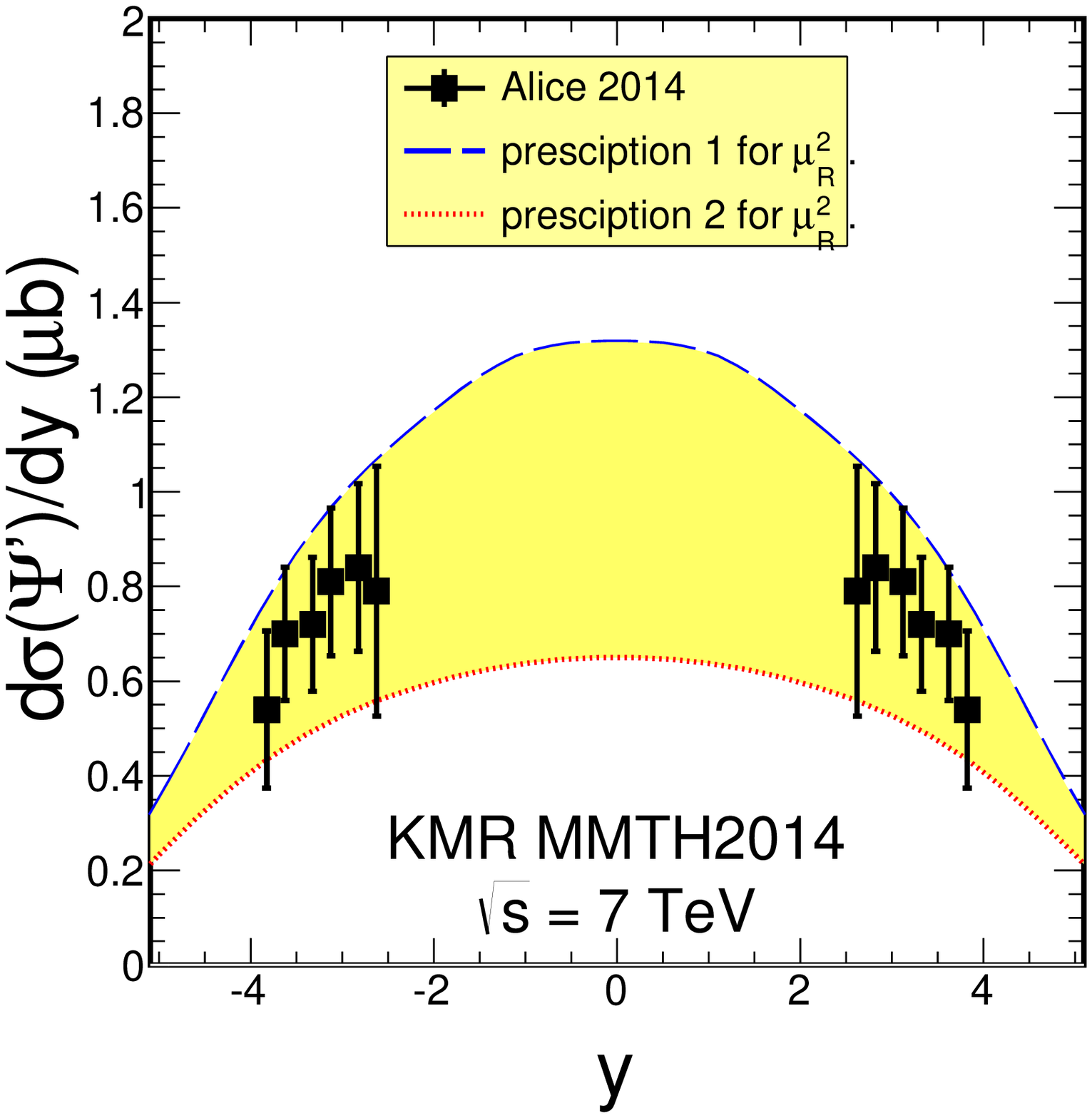}
\includegraphics[width=5.5cm]{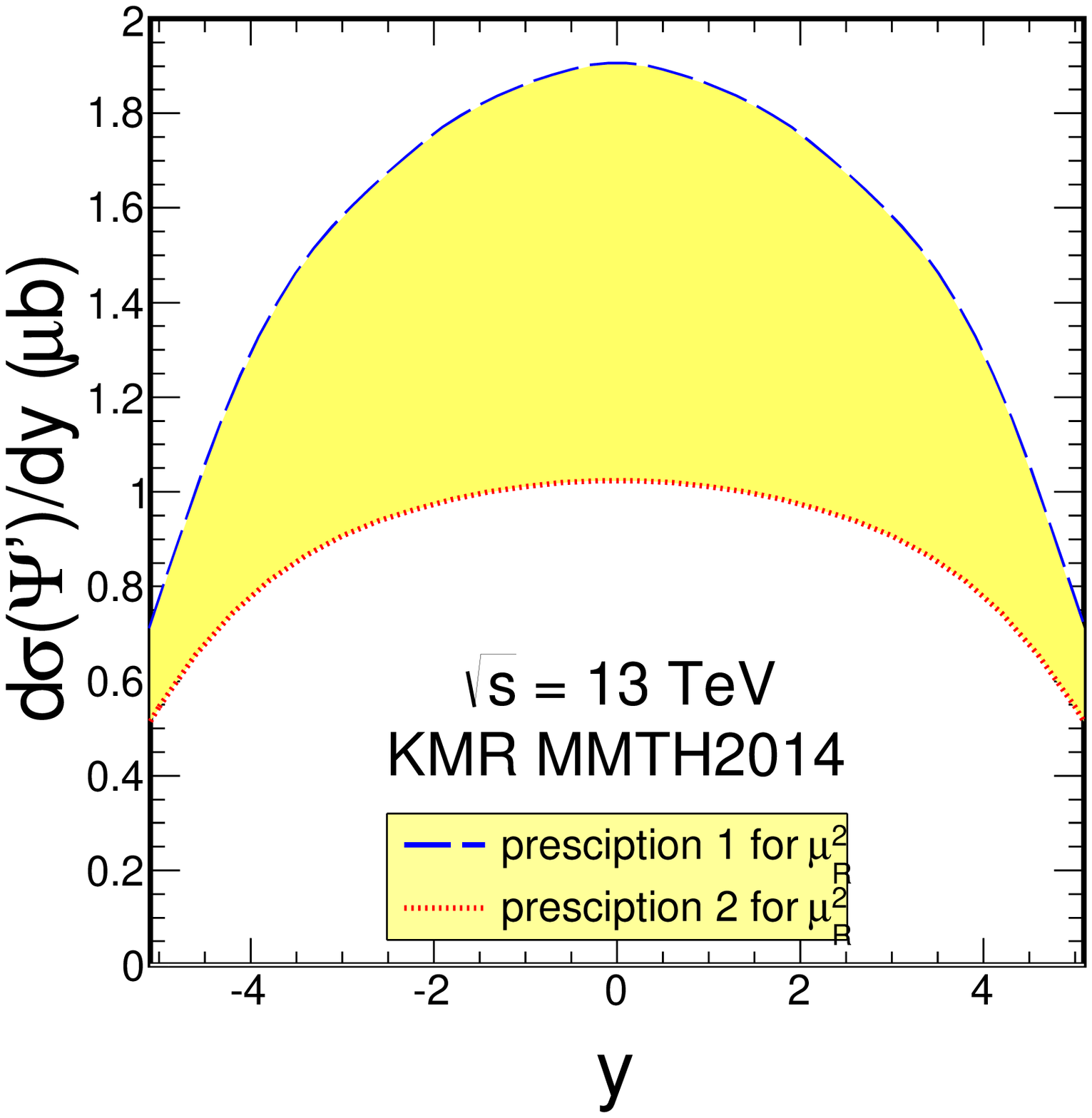}
\caption{
Rapidity distribution of $\psi'$ meson (direct mechanism)
for the KMR UGDF for $\sqrt{s} =$ 7 TeV (left panel)
and $\sqrt{s} =$ 13 TeV (right panel).
The upper line is for the scale prescription 1 and the lower line is
for prescription 2.
}
\label{fig:dsig_dy_psip_KMR}
\end{figure}

We start with $\psi'$ production.
In Fig.\ref{fig:dsig_dy_psip_KMR} we show rapidity distributions
of $\psi'$ obtained with the KMR unintegrated gluon distributions.
In the left panel we compare our results with the ALICE experimental
data for $\sqrt{s}$ = 7 TeV \cite{ALICE_2014_7TeV}. The ALICE collaboration measured only
$\psi'$ mesons emitted in rather forward directions. This corresponds to
one longitudinal momentum fraction small and the second longitudinal momentum
fraction large.
We show results with the two different prescriptions for the arguments
of QCD running coupling constant as was discussed in the previous section.
One gets rather large uncertainty band associated with the choice
of the $\alpha_s$ argument. On the right panel we show our predictions for 
$\sqrt{s}$ = 13 TeV.

\begin{figure}
\includegraphics[width=5.5cm]{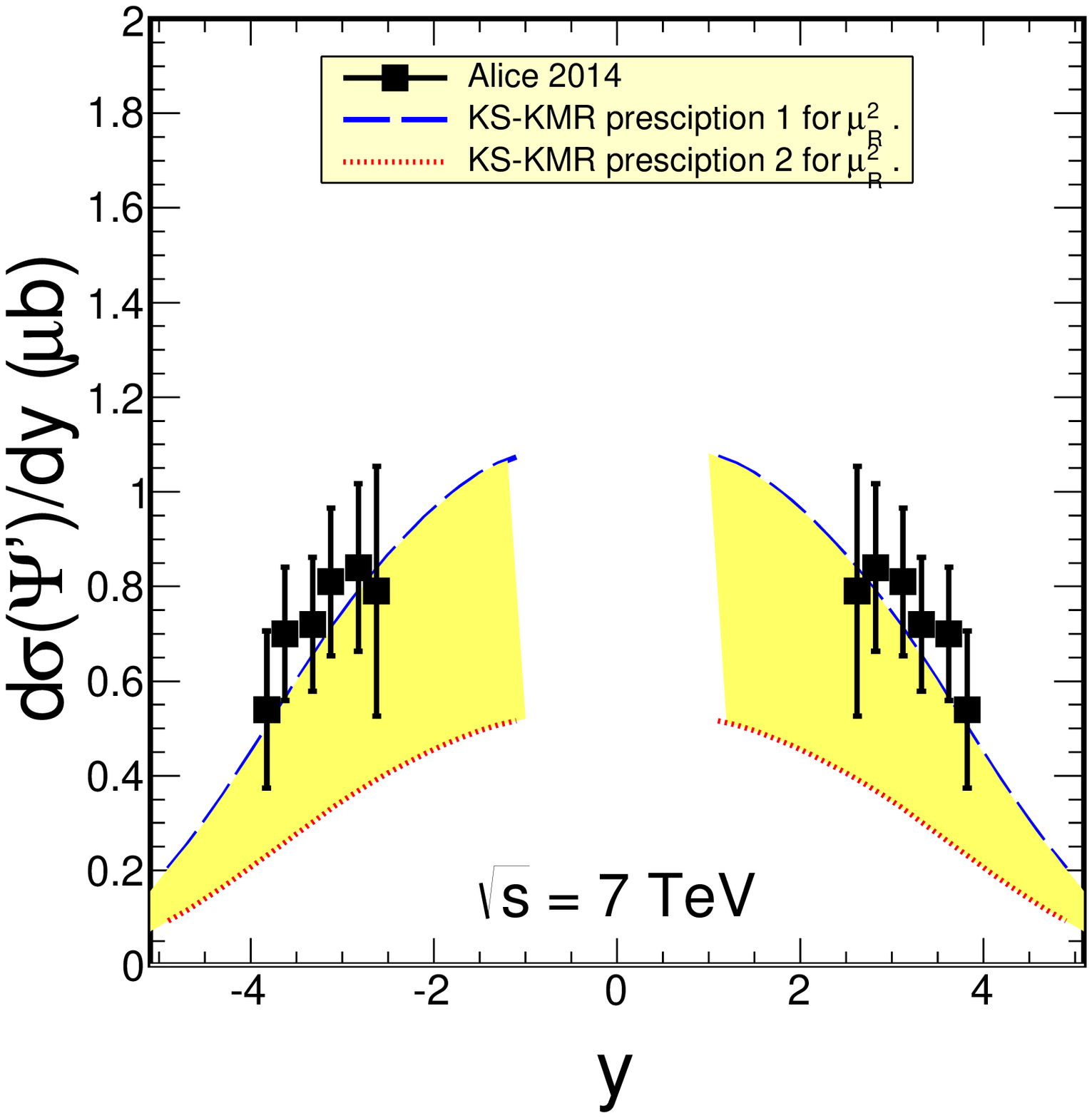}
\includegraphics[width=5.5cm]{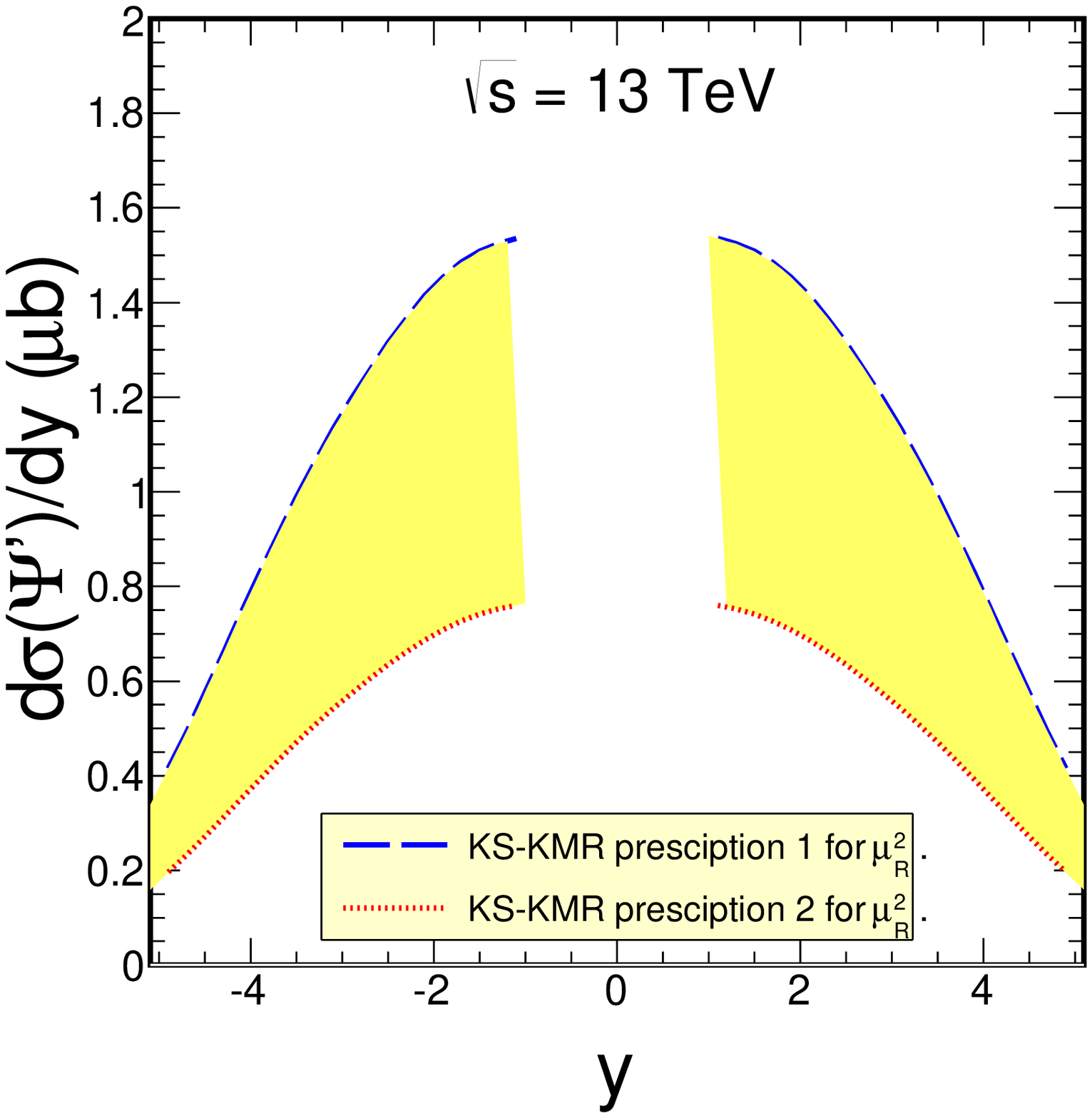}
\caption{
Rapidity distribution of $\psi'$ meson (direct mechanism)
for the mixed UGDFs for $\sqrt{s} =$ 7 TeV (left panel)
and $\sqrt{s} =$ 13 TeV (right panel).
The upper line is for scale prescription 1 and the lower line is
for prescription 2.
}
\label{fig:dsig_dy_psip_mixed}
\end{figure}

For very small $x$ one may expect saturation effects. The KMR UGDF
does not include such effects. To illustrate potential effect instead
of using both UGDFs of the same type (KMR) for large $x_1/x_2$ we take
the KMR UGDF, whereas for small $x_2/x_1$ we take for example the
Kutak-Sta\'sto nonlinear UGDF \cite{KS2005}. It is marked by KS acronym
in Fig.\ref{fig:dsig_dy_psip_mixed}. Slightly smaller cross section
has been obtained than with the KMR UGDF. The effect is, however, not significant.

Since the $\psi'$ meson decays: $\psi' \to J/\psi X$ with BF = 0.61
\cite{PDG} it constitutes also a contribution to the $J/\psi$ channel
and will be taken into account in the rest of the paper.

\subsection{$J/\Psi$ direct production}

\begin{figure}
\includegraphics[width=5.0cm]{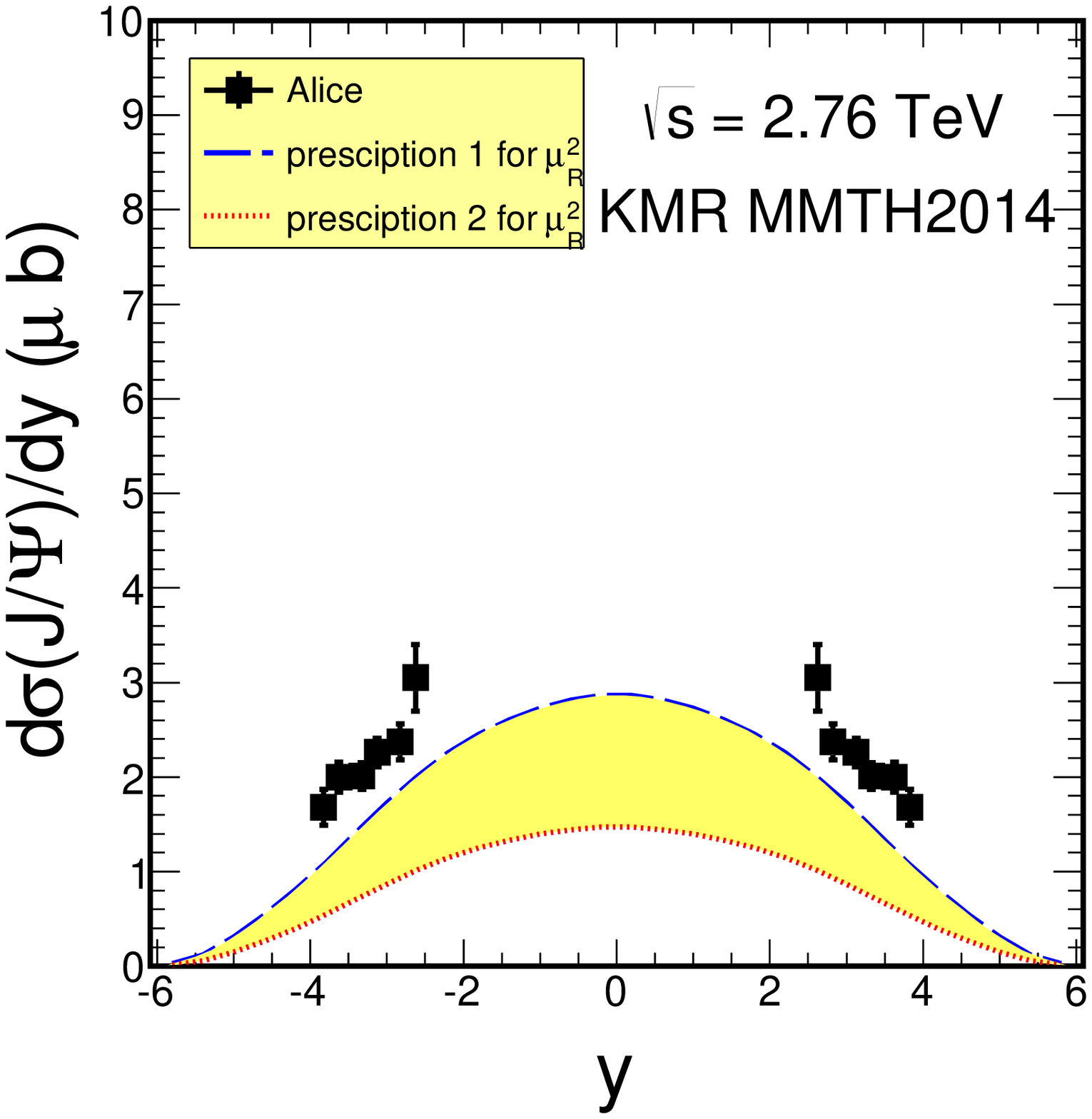}
\includegraphics[width=5.0cm]{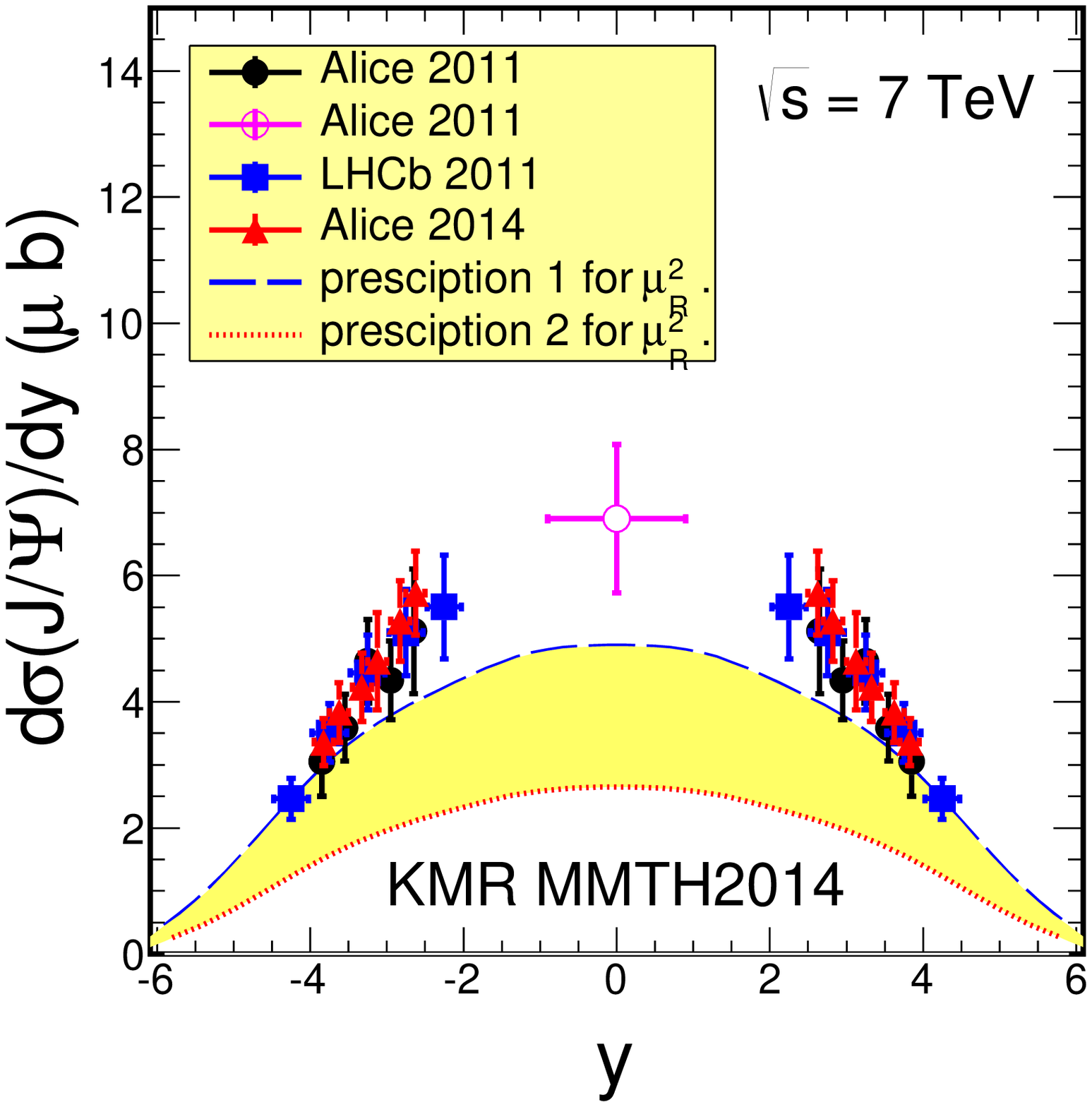}
\includegraphics[width=5.0cm]{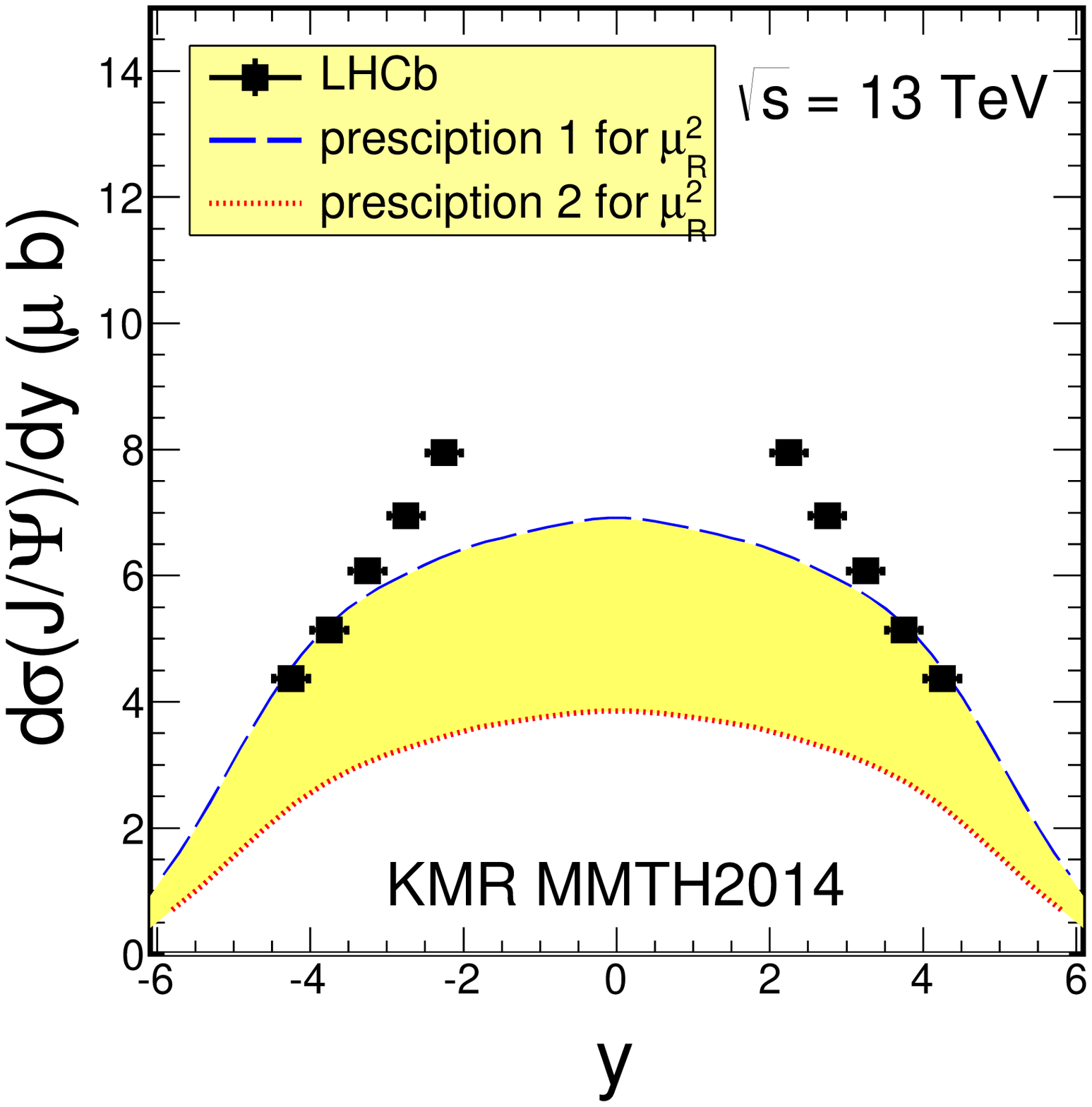}
\caption{
Rapidity distribution of $J/\psi$ mesons (direct mechanism)
for the KMR UGDF for $\sqrt{s} =$ 2.76 TeV (left panel), 
$\sqrt{s} =$ 7 TeV (middle panel)
and $\sqrt{s} =$ 13 TeV (right panel).
The upper line is for the scale prescription 1 and the lower line is
for prescription 2. The results are compared with the ALICE \cite{ALICE_2012_2760,ALICE_2011_7TeV,
ALICE_2014_7TeV} and LHCb \cite{LHCb_2011_7TeV,LHCb_2015_13TeV} experimental data.
}
\label{fig:dsig_dy_jpsi_direct_KMR}
\end{figure}

\begin{figure}
\includegraphics[width=5.0cm]{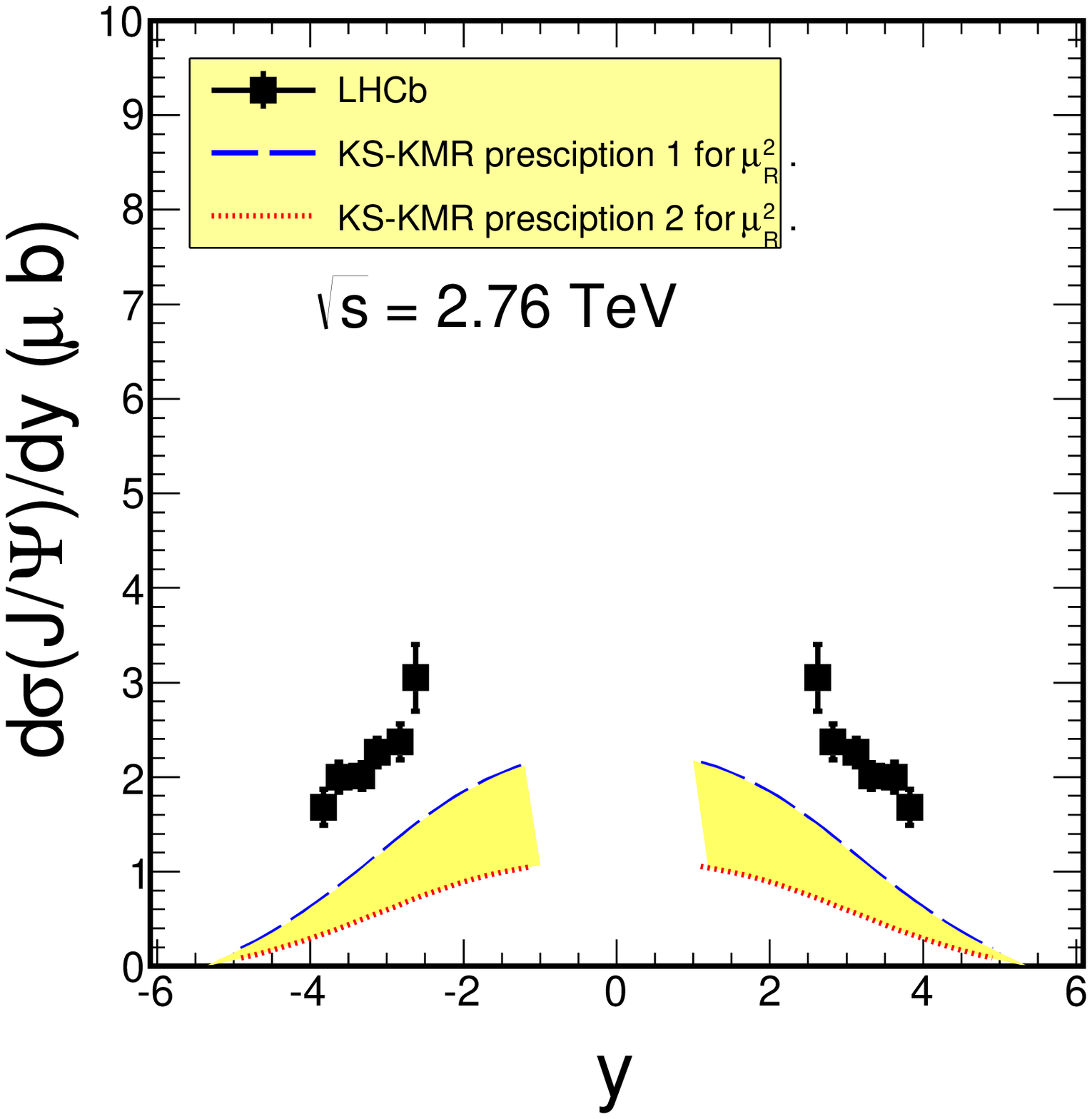}
\includegraphics[width=5.0cm]{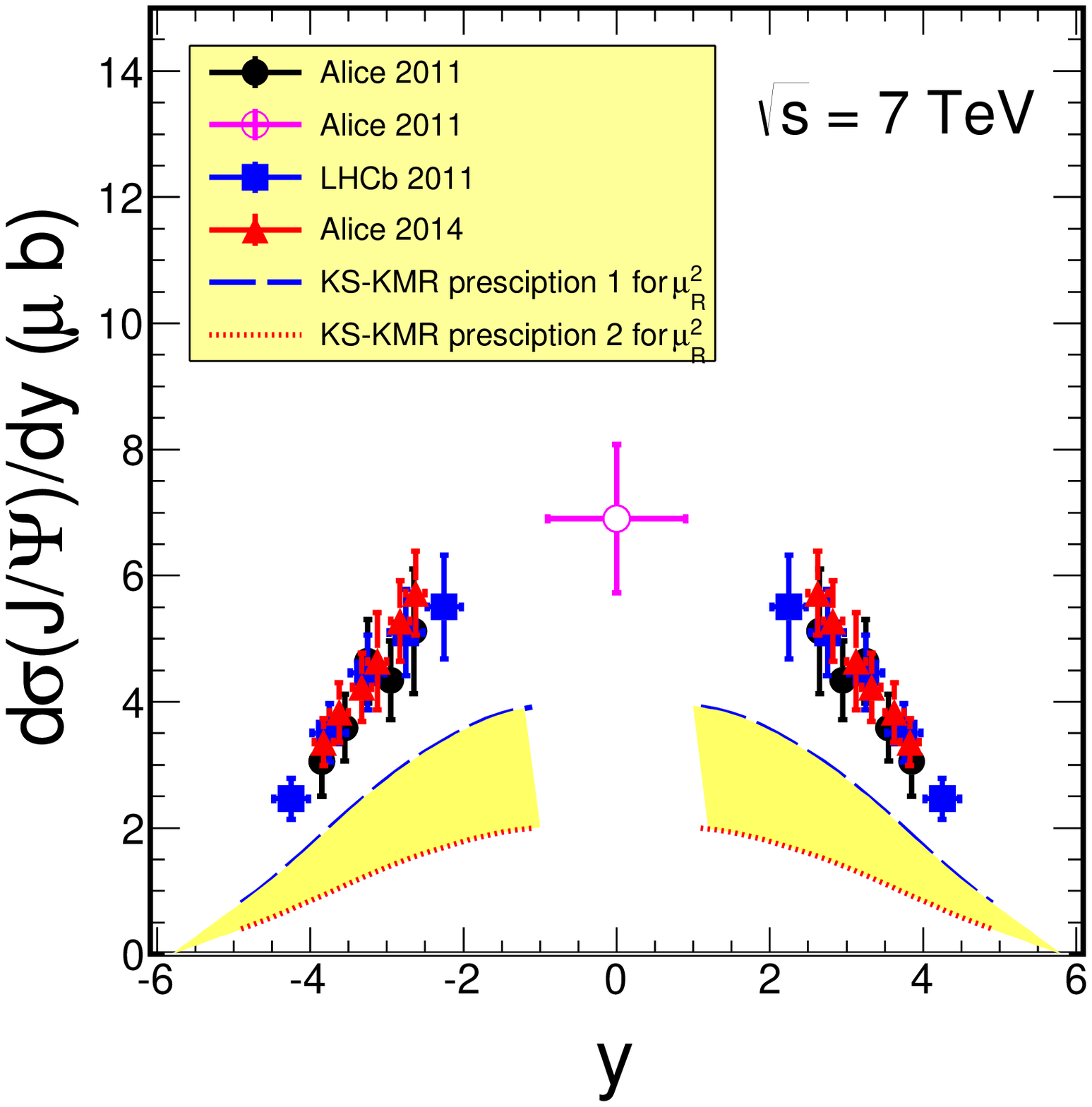}
\includegraphics[width=5.0cm]{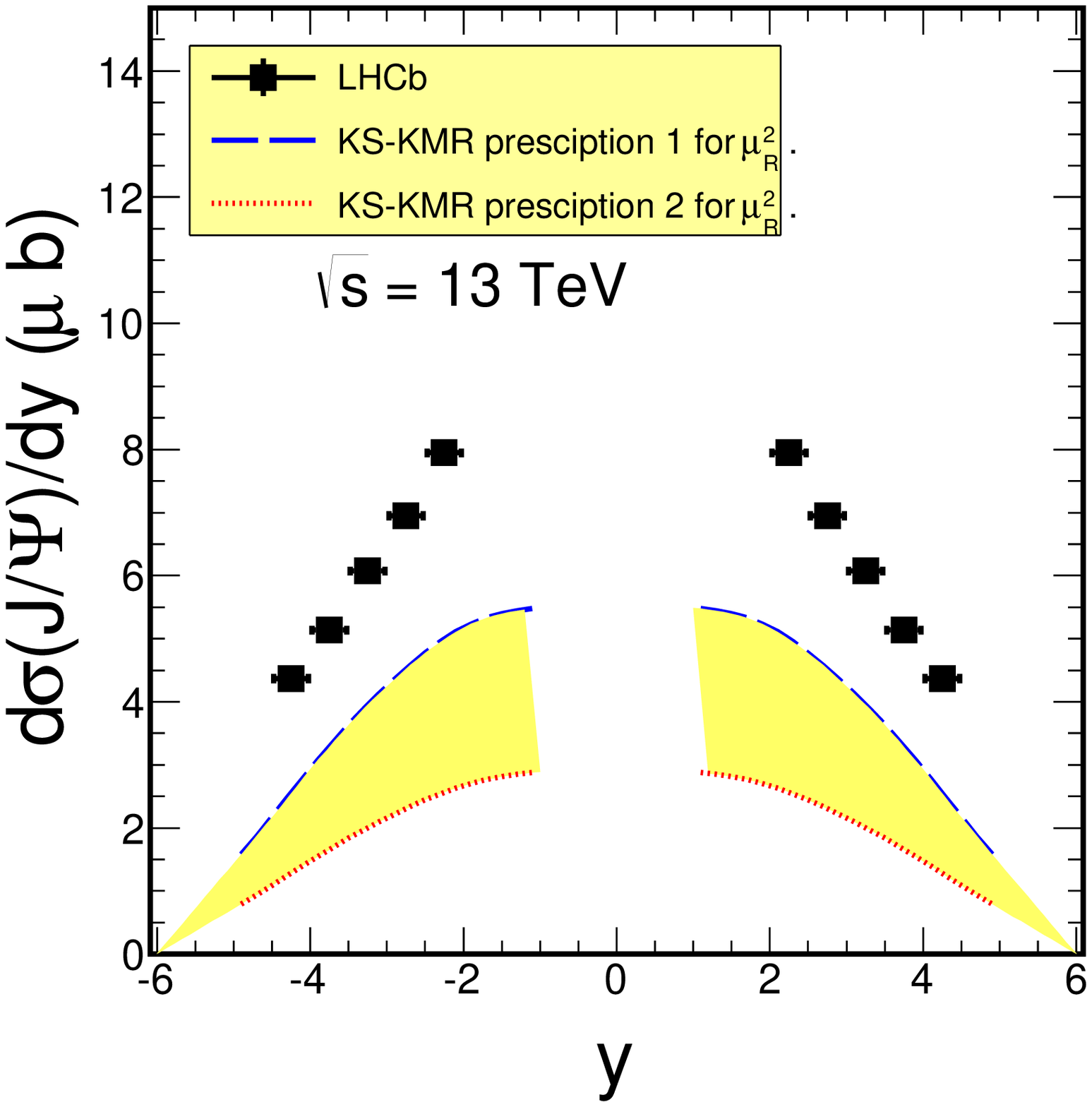}
\caption{
Rapidity distribution of $J/\psi$ mesons (direct mechanism)
for the mixed UGDFs for $\sqrt{s} =$ 2.76 GeV (left panel), 
$\sqrt{s} =$ 7 TeV (middle panel) and $\sqrt{s} =$ 13 TeV (right panel).
The upper line is for the scale prescription 1 and the lower line is
for prescription 2.
}
\label{fig:dsig_dy_jpsi_direct_mixed}
\end{figure}

\begin{figure}
\includegraphics[width=5.5cm]{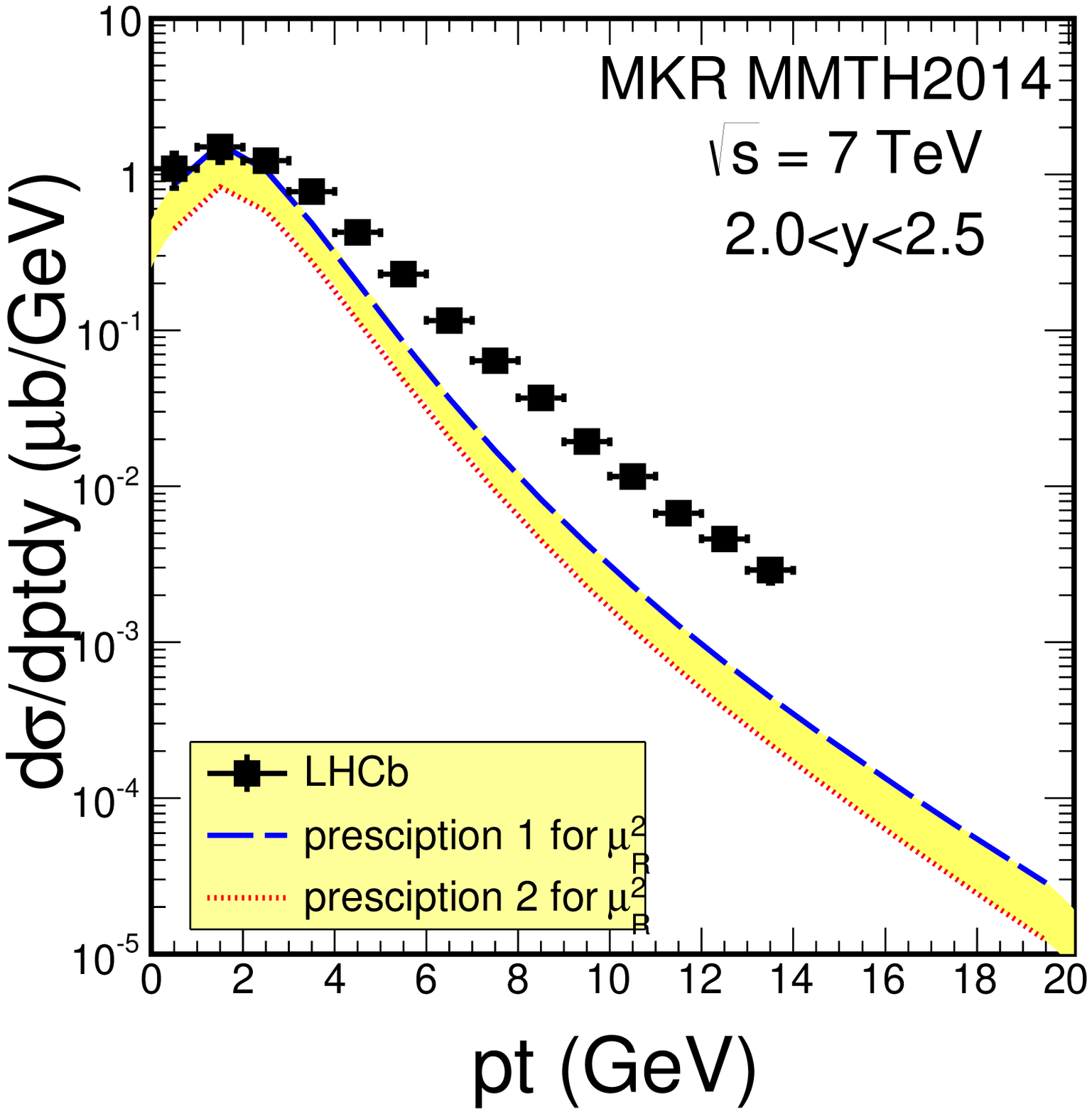}
\includegraphics[width=5.5cm]{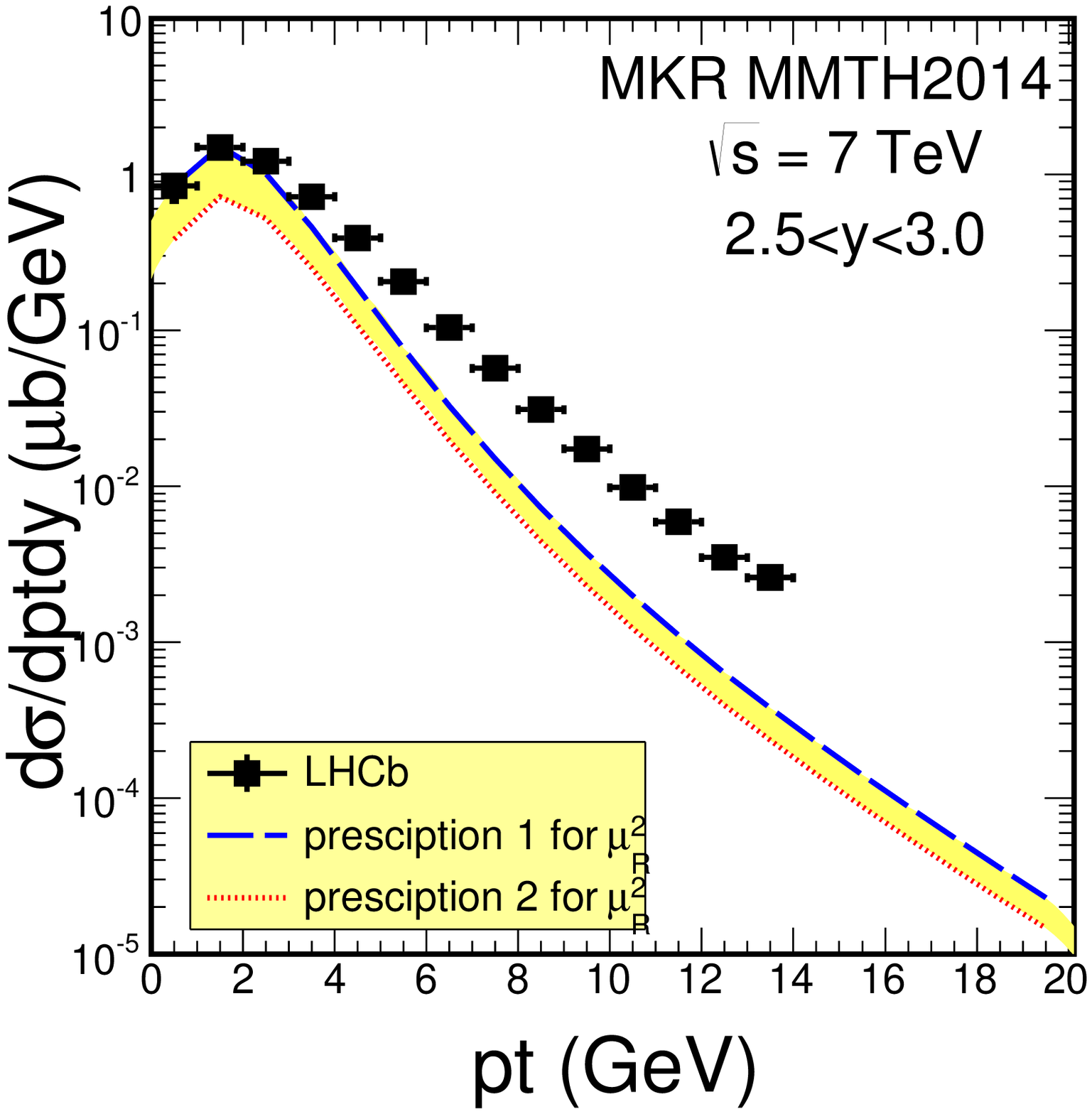}\\
\includegraphics[width=5.5cm]{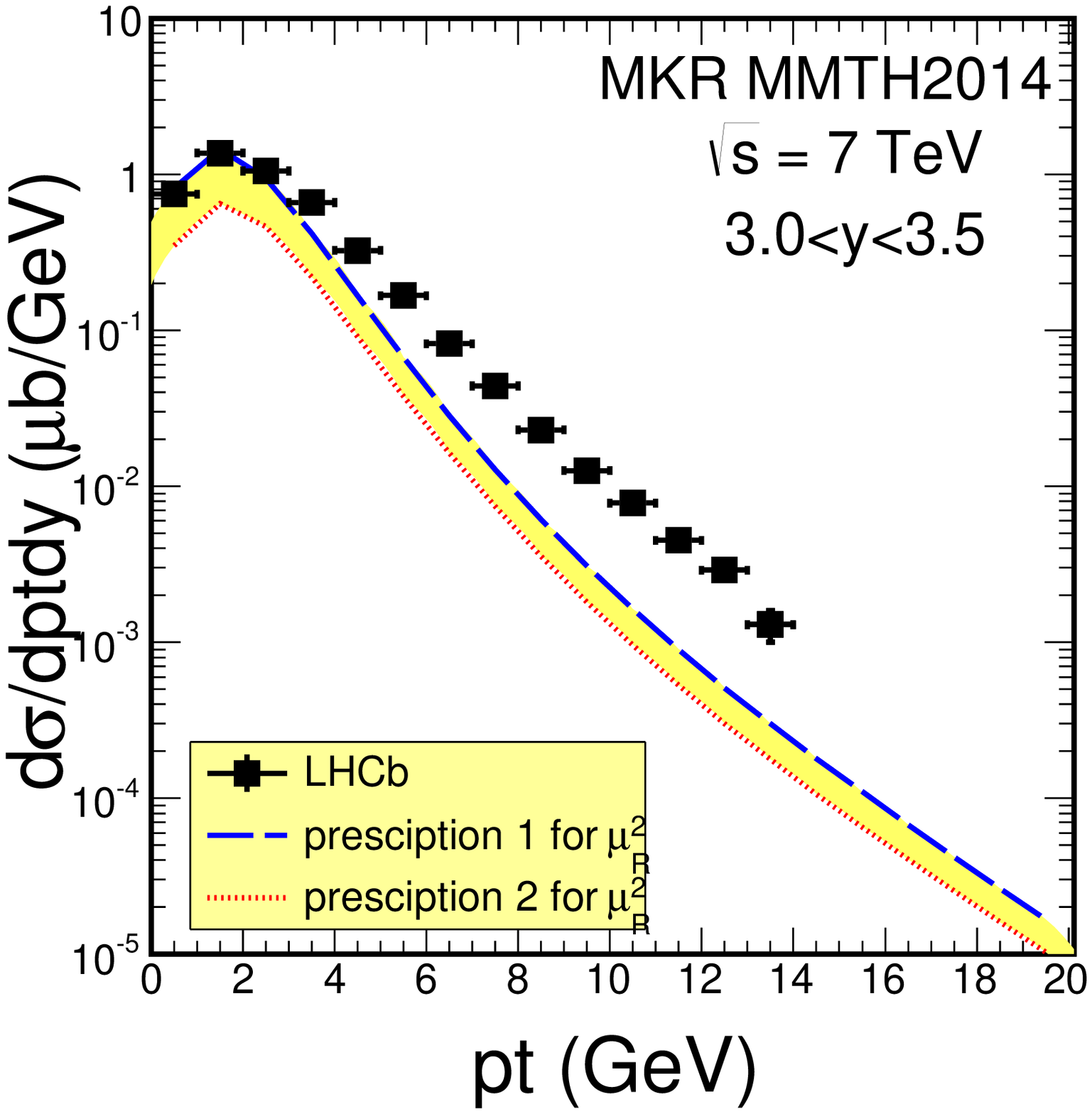}
\includegraphics[width=5.5cm]{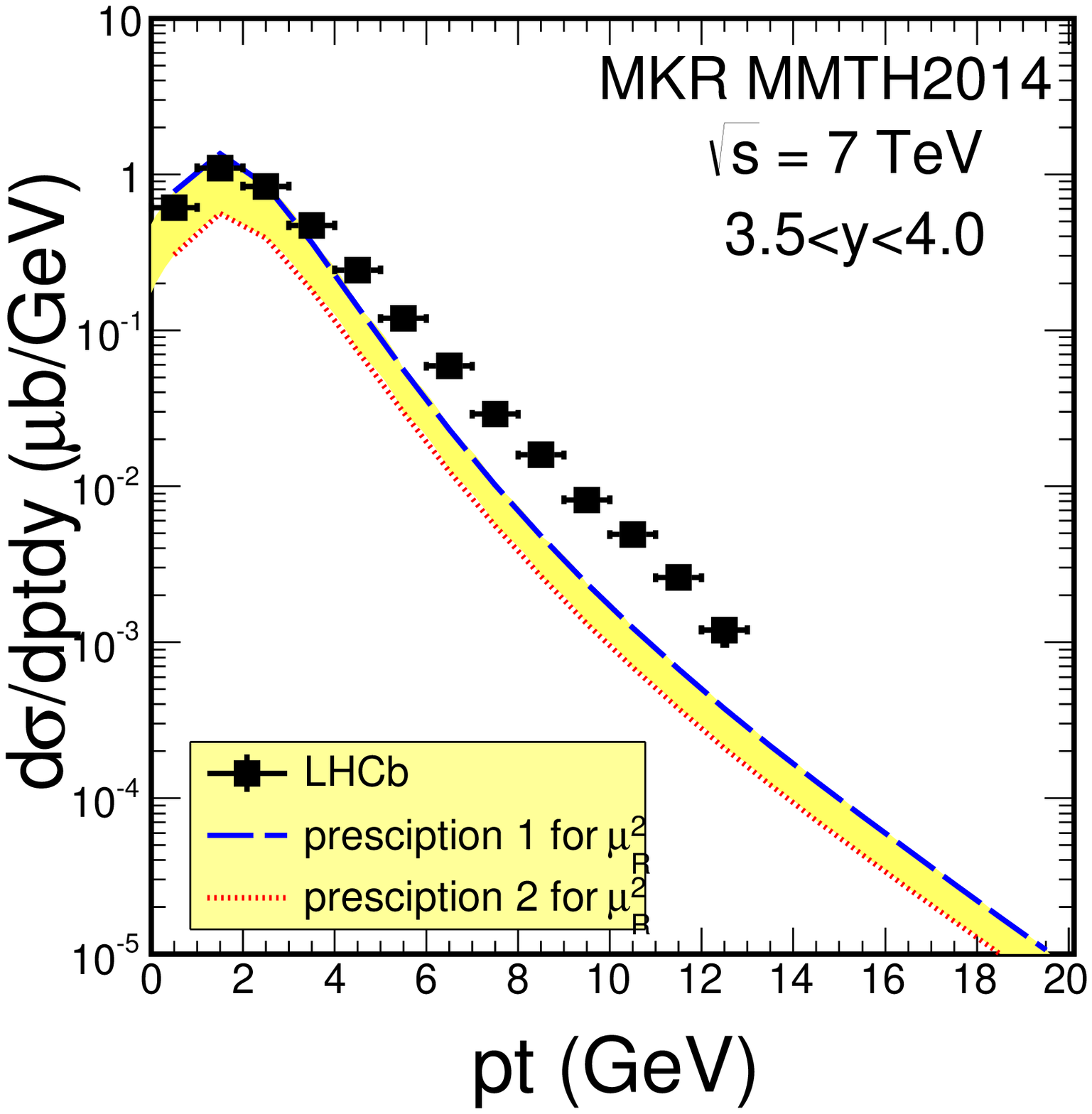}
\caption{
Transverse momentum distribution of $J/\psi$ (direct componet only)
together with LHCb \cite{LHCb_2011_7TeV} experimental data for $\sqrt{s} =$ 7 TeV for 
different ranges of rapidity specified in the figures.}
\label{fig:dsig_dpt_jpsi_KMR_7TeV}
\end{figure}

\begin{figure}
\includegraphics[width=5.5cm]{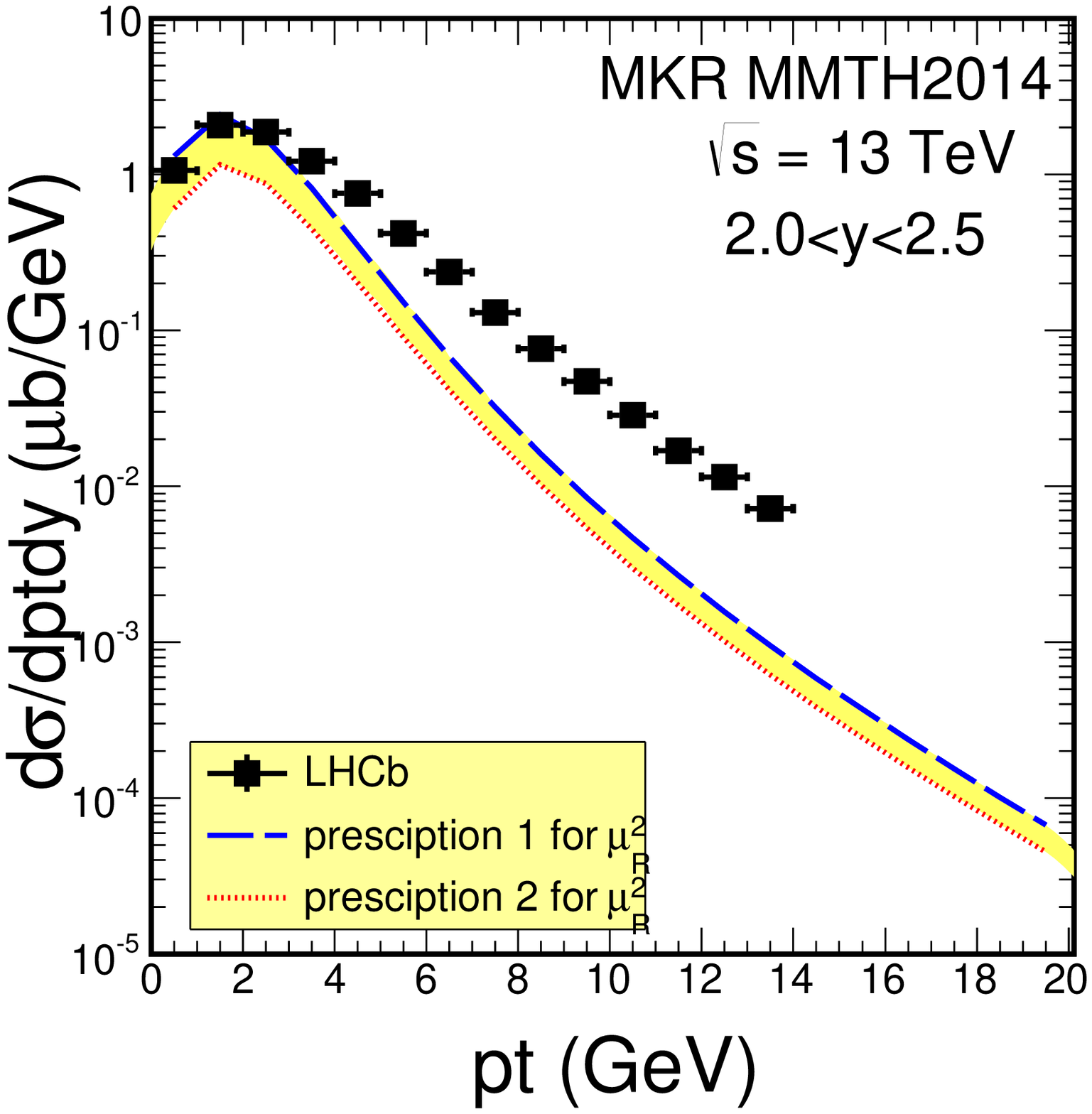}
\includegraphics[width=5.5cm]{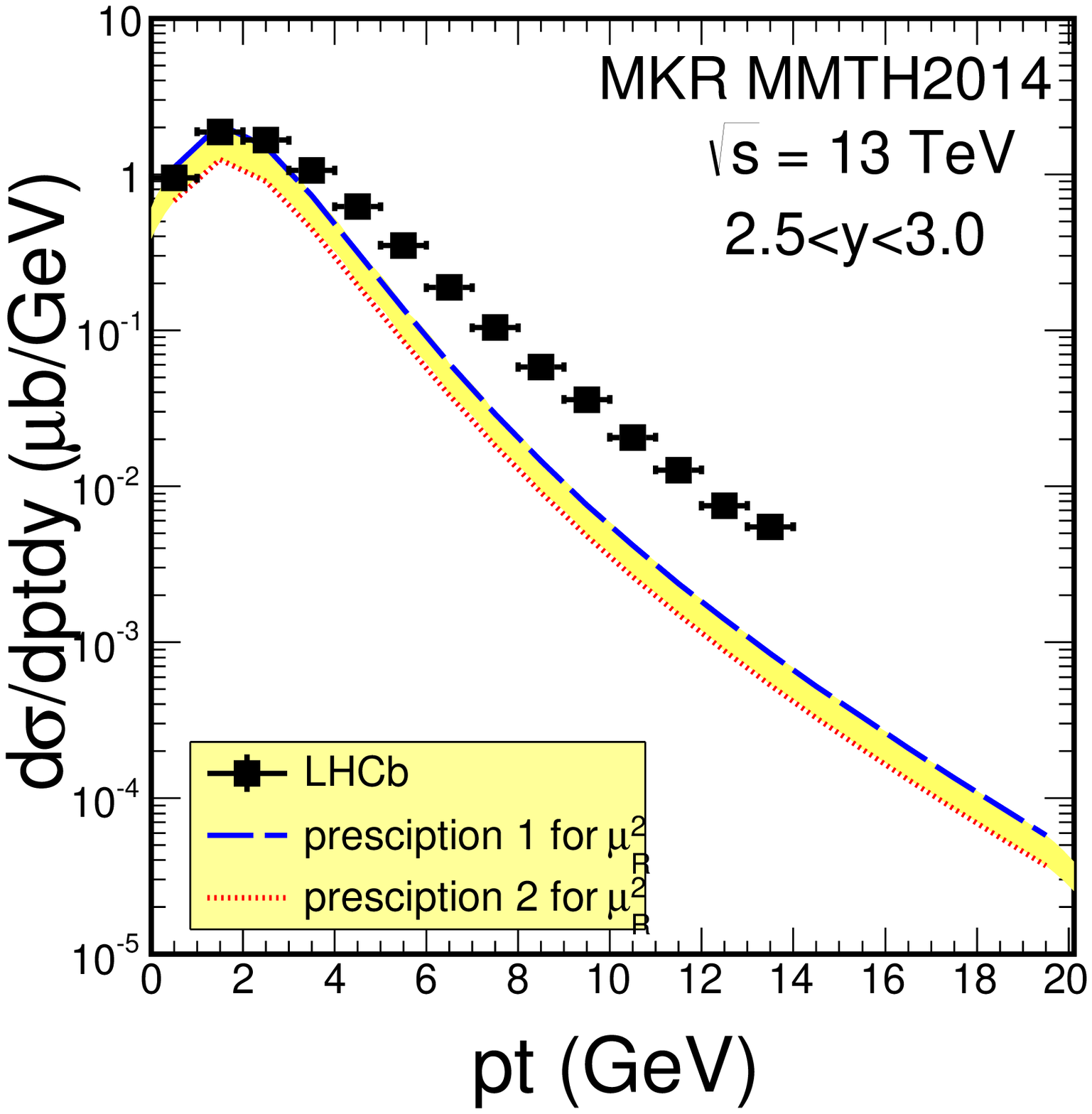}\\
\includegraphics[width=5.5cm]{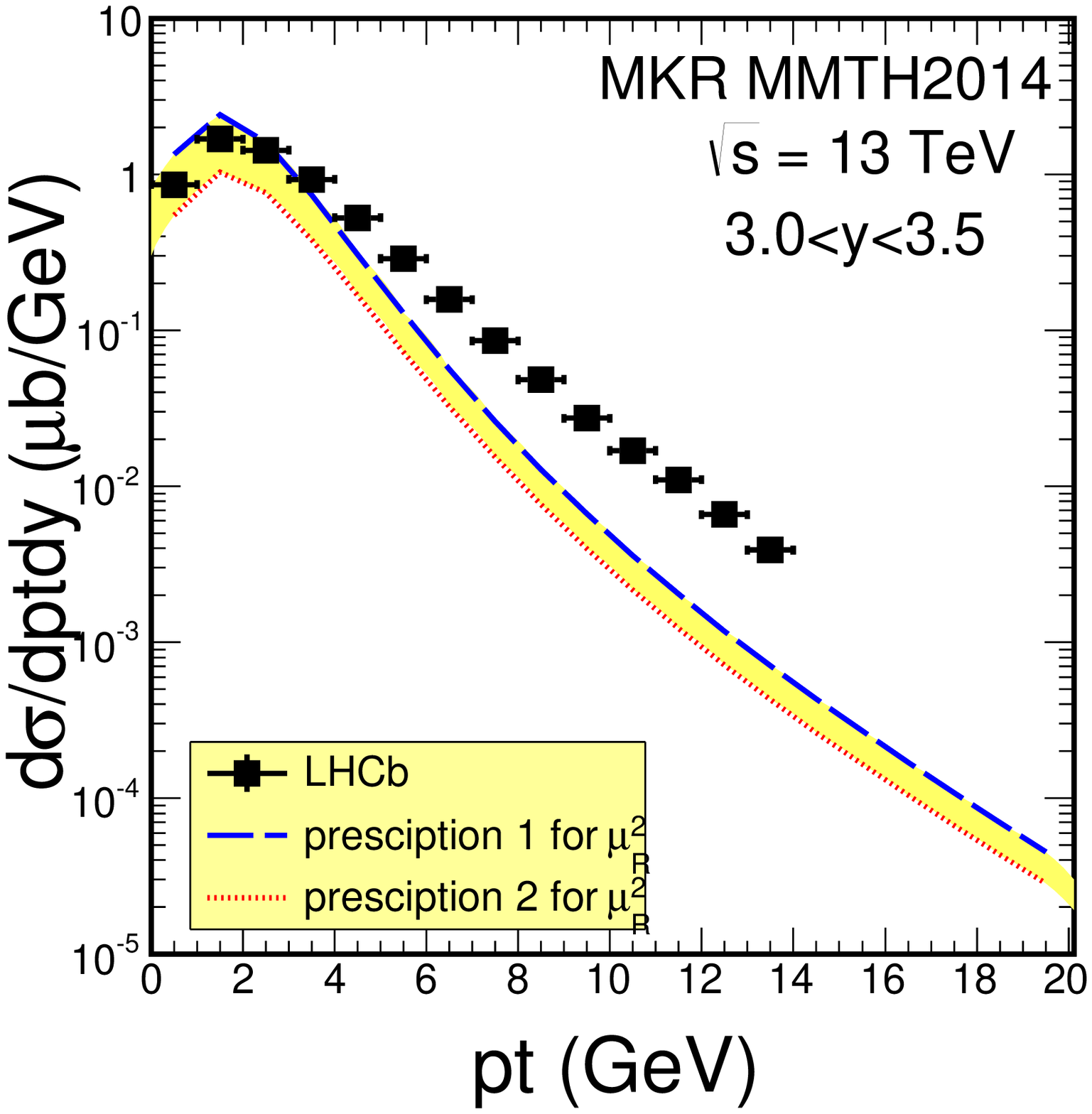}
\includegraphics[width=5.5cm]{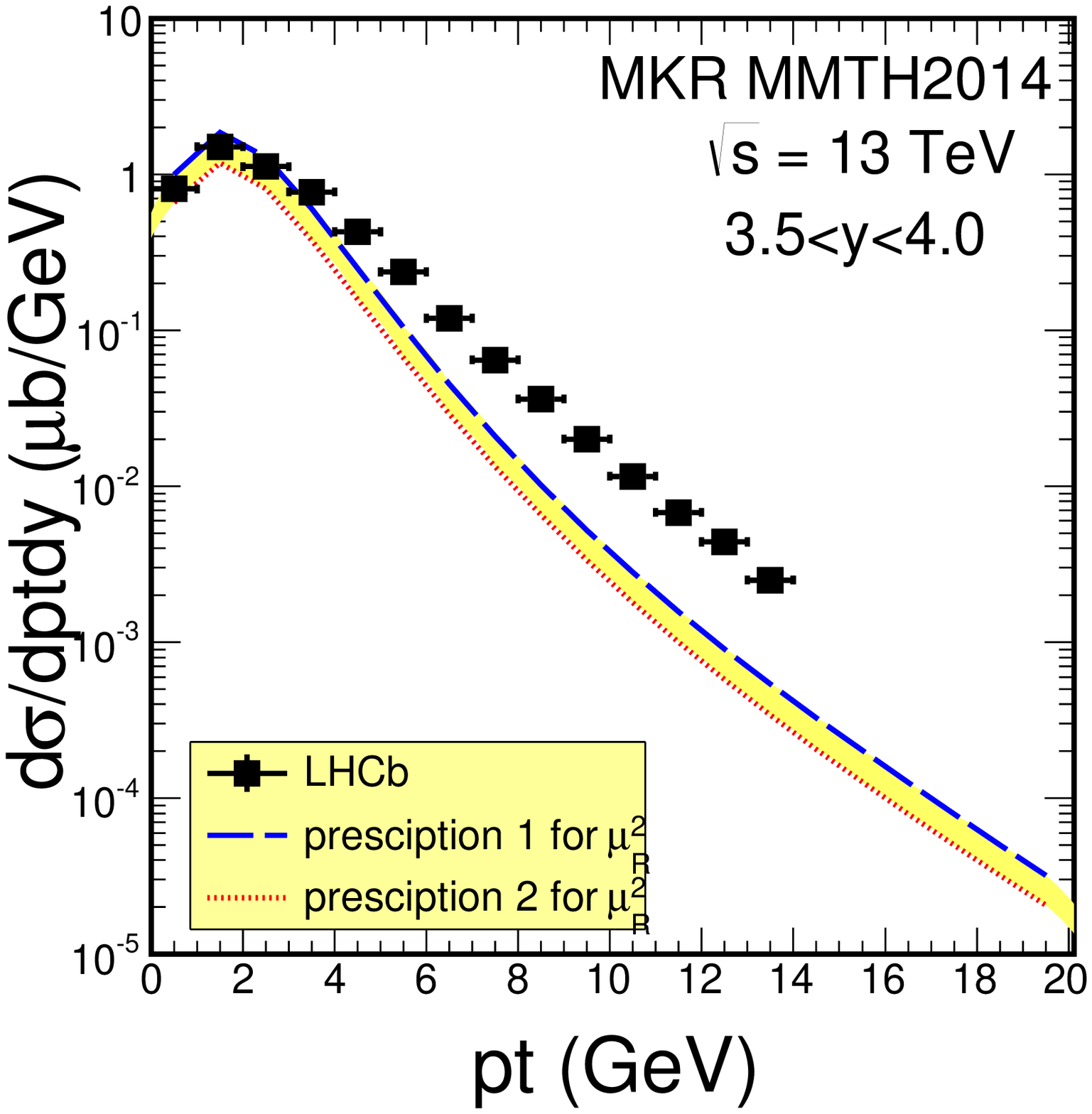}
\caption{
Transverse momentum distribution of $J/\psi$ (direct componet only)
together with LHCb \cite{LHCb_2015_13TeV} experimental data for $\sqrt{s} =$ 13 TeV for 
different ranges of rapidity specified in the figures.}
\label{fig:dsig_dpt_jpsi_KMR_13TeV}
\end{figure}

There are three components of prompt $J/\psi$ production:
direct production (see Fig. \ref{fig:dsig_dy_jpsi_direct_KMR}-
\ref{fig:dsig_dpt_jpsi_KMR_13TeV}), and feed down from
$\psi'$ and $\chi_c$ decays. 

In this subsection we present results for the direct component for
$J/\psi$ production. In Fig.\ref{fig:dsig_dy_jpsi_direct_KMR} we show
exclusively this contribution for three different collision energies:
$\sqrt{s}$ = 2.76 TeV (left panel), $\sqrt{s}$ = 7 TeV (middle panel)
and $\sqrt{s}$ = 13 TeV (right panel). As for $\psi'$ production we
show our results for two different prescriptions for $\alpha_s$
and for the KMR UGDF. 
As for the $\psi'$ production there is large uncertainty related to the choice
of running coupling constant (see the yellow band).
The direct contribution is large but there
is a room for other contributions, to be discussed in the following.

For completeness in Fig.\ref{fig:dsig_dy_jpsi_direct_mixed}
we show also our results with the mixed UGDFs scenario described above.
Here the effect of UGDF modification is similar as for $\psi'$.

We shall look also for transverse momentum distributions.
For example the LHCb collaboration measured such distributions
for different intervals of rapidity \cite{LHCb_2011_7TeV,LHCb_2015_13TeV}.
In Fig.\ref{fig:dsig_dpt_jpsi_KMR_7TeV} we show such distributions
for $\sqrt{s}$ = 7 TeV for two different prescriptions of $\alpha_s$. 
Our direct component exhausts large fraction of the cross section 
for small $p_t$. At larger $p_t$ clearly some other
contributions are missing.

In Fig.\ref{fig:dsig_dpt_jpsi_KMR_13TeV} we show similar
distributions for $\sqrt{s}$ = 13 TeV. The situation is very much the
same as for $\sqrt{s}$ = 7 TeV.

\subsection{$J/\psi$ from $\chi_c$ decays}

\begin{figure}
\includegraphics[width=5.0cm]{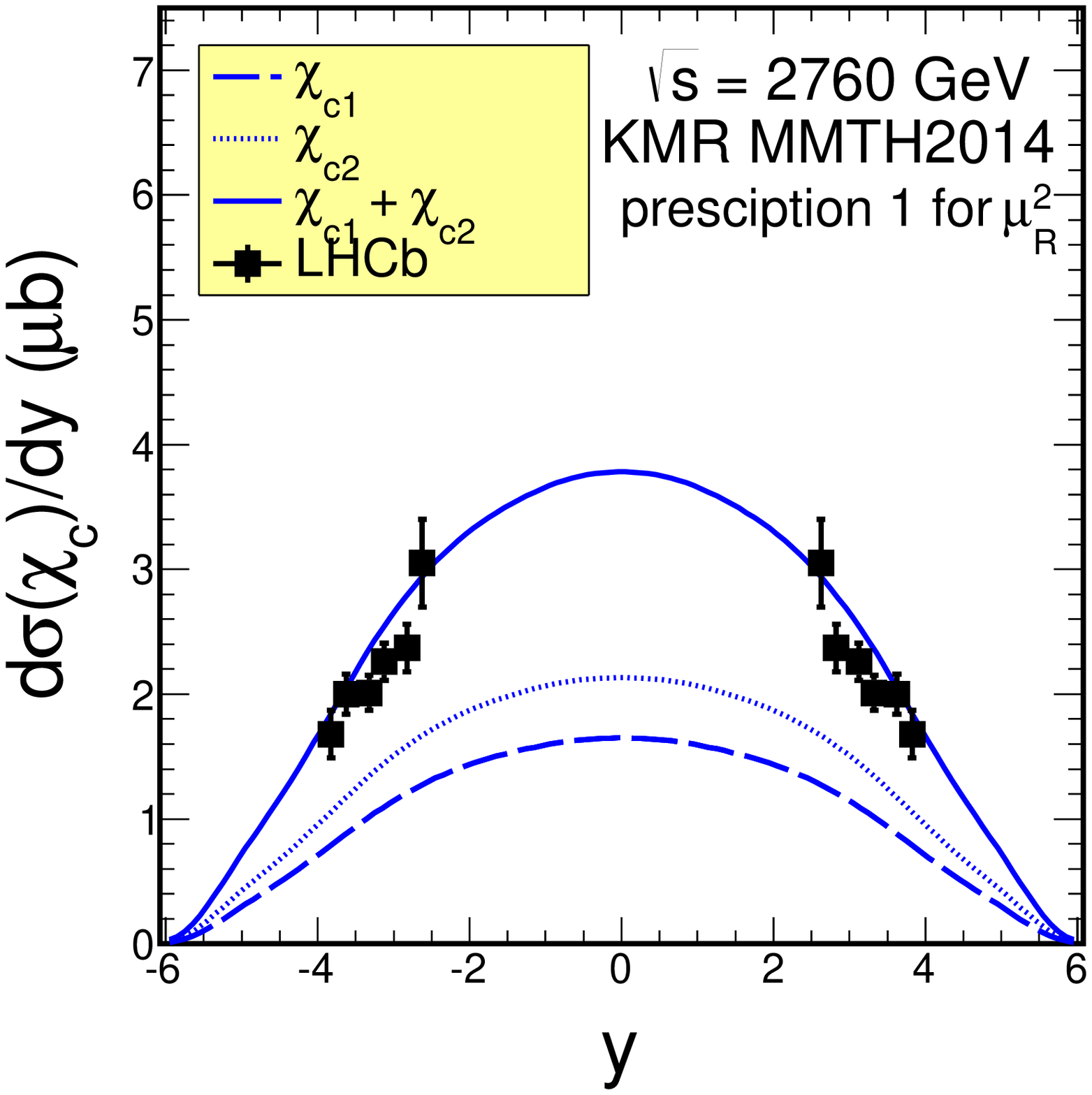}
\includegraphics[width=5.0cm]{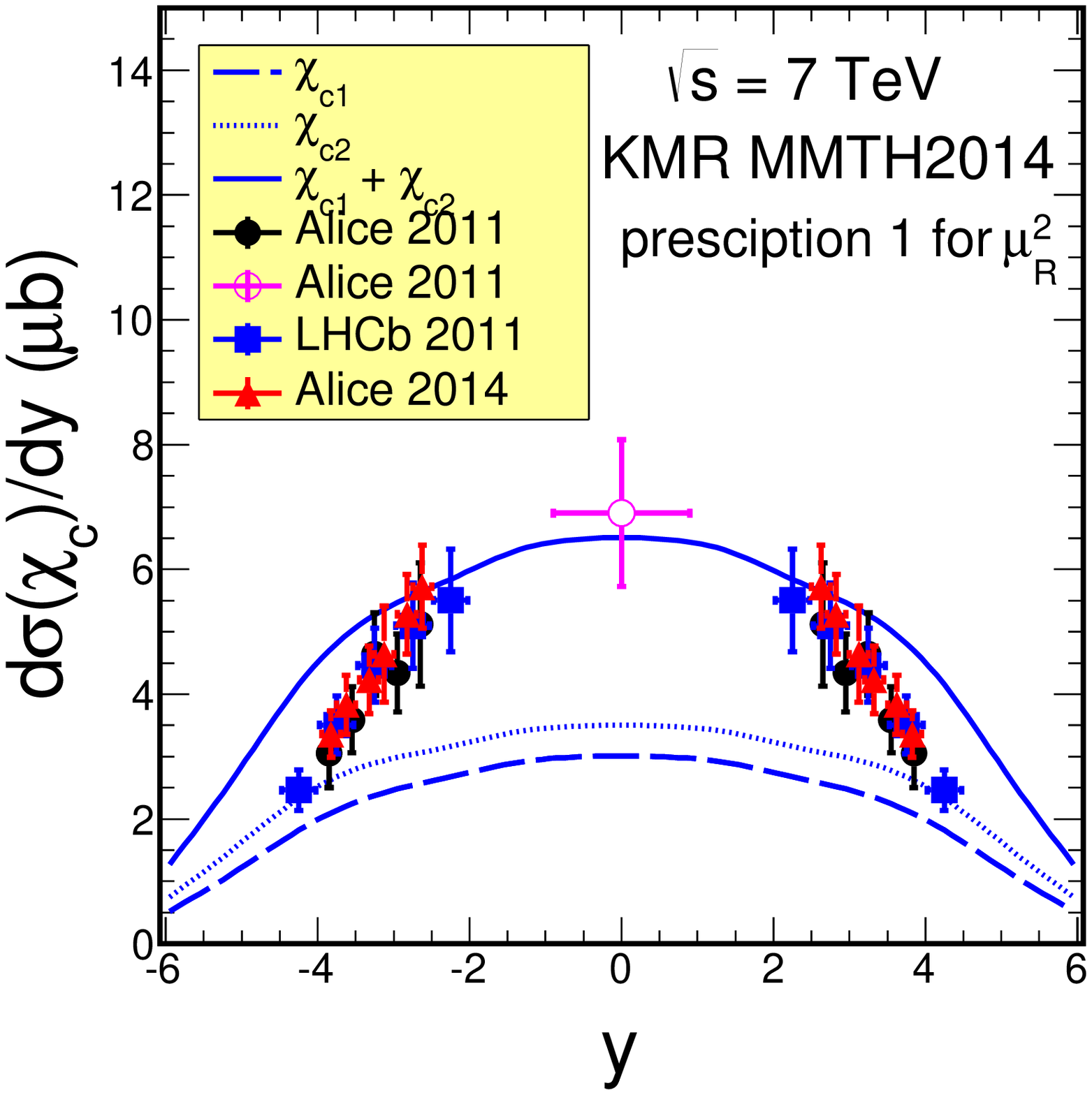}
\includegraphics[width=5.0cm]{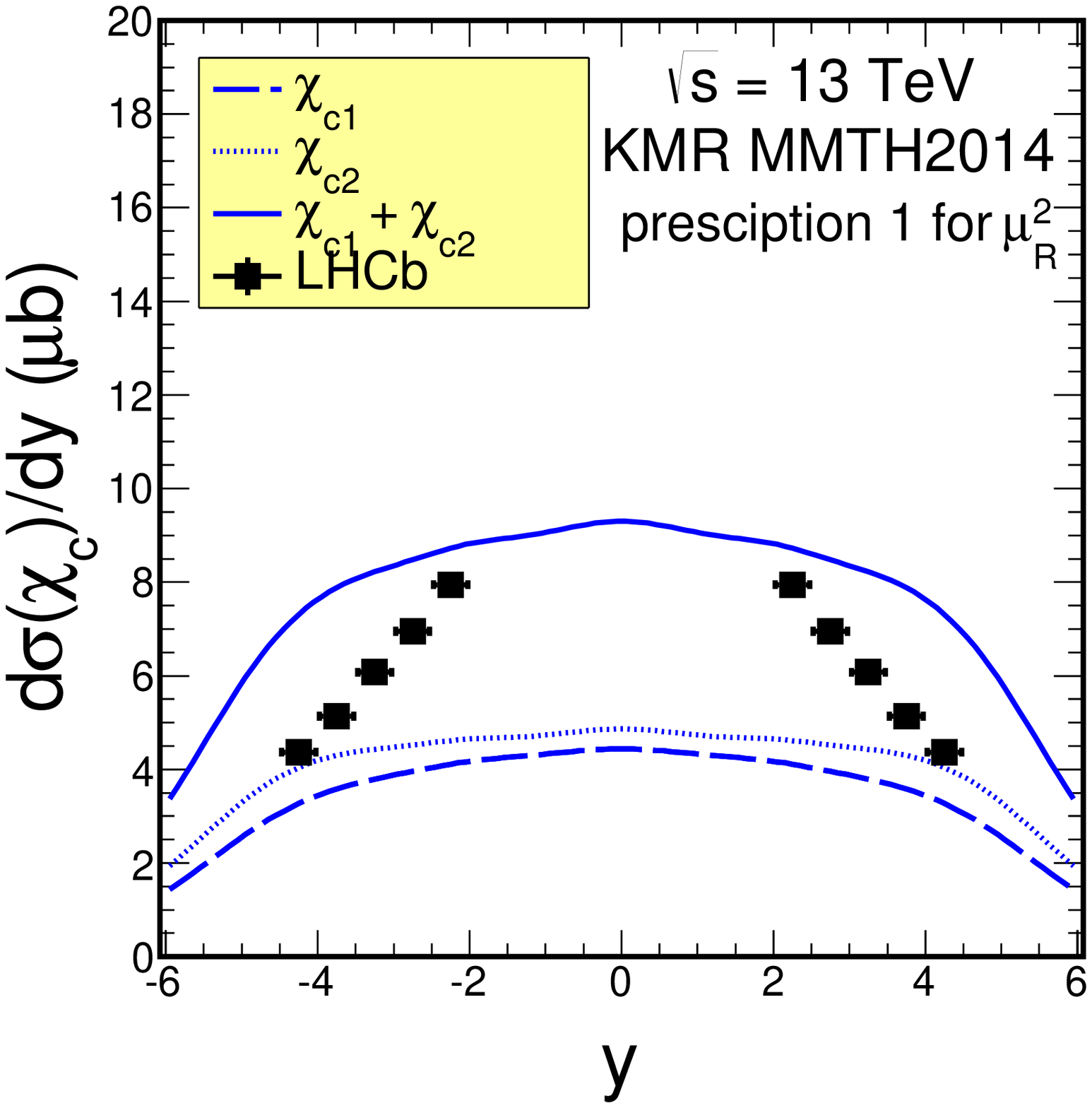} \\
\includegraphics[width=5.0cm]{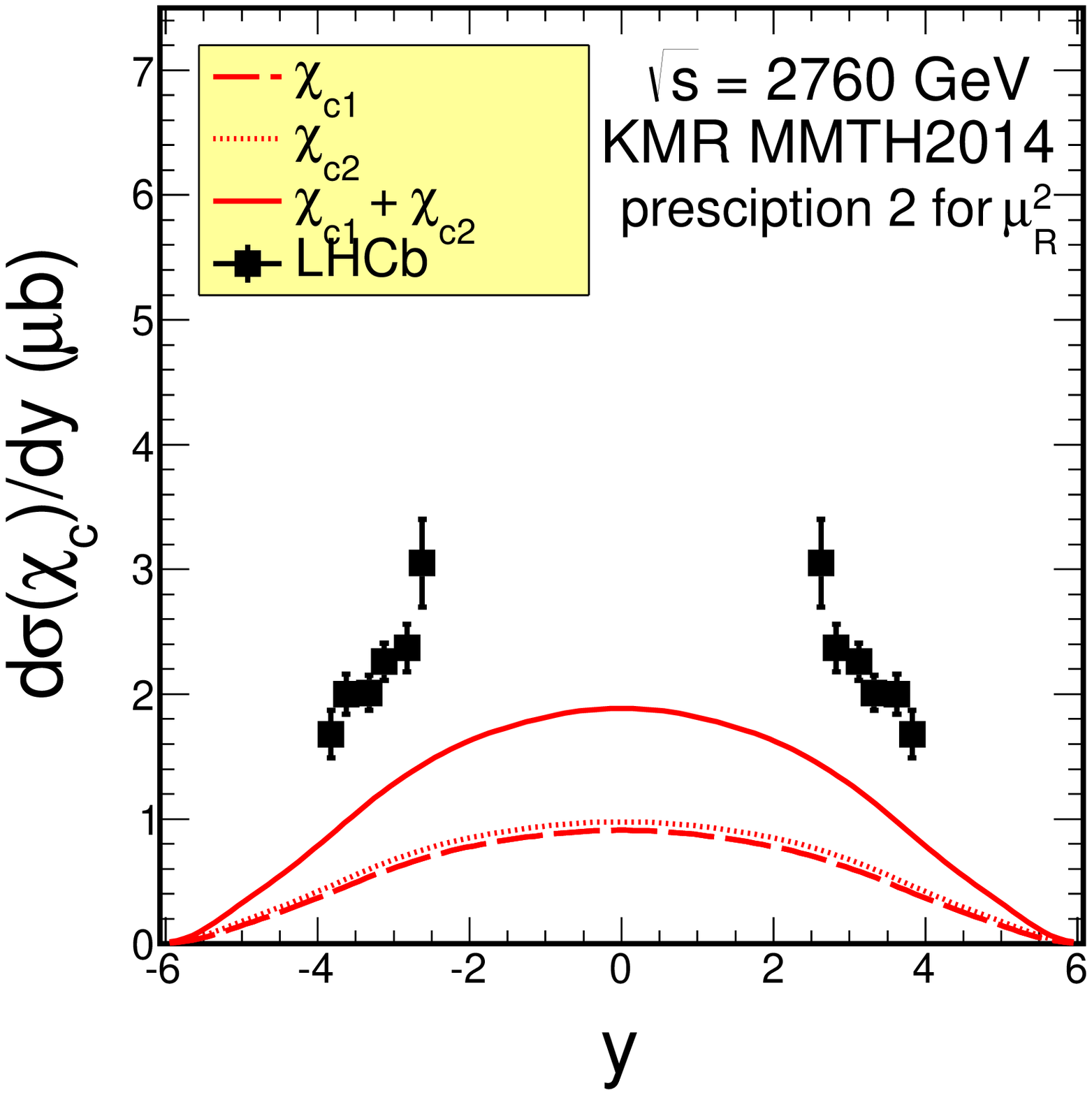}
\includegraphics[width=5.0cm]{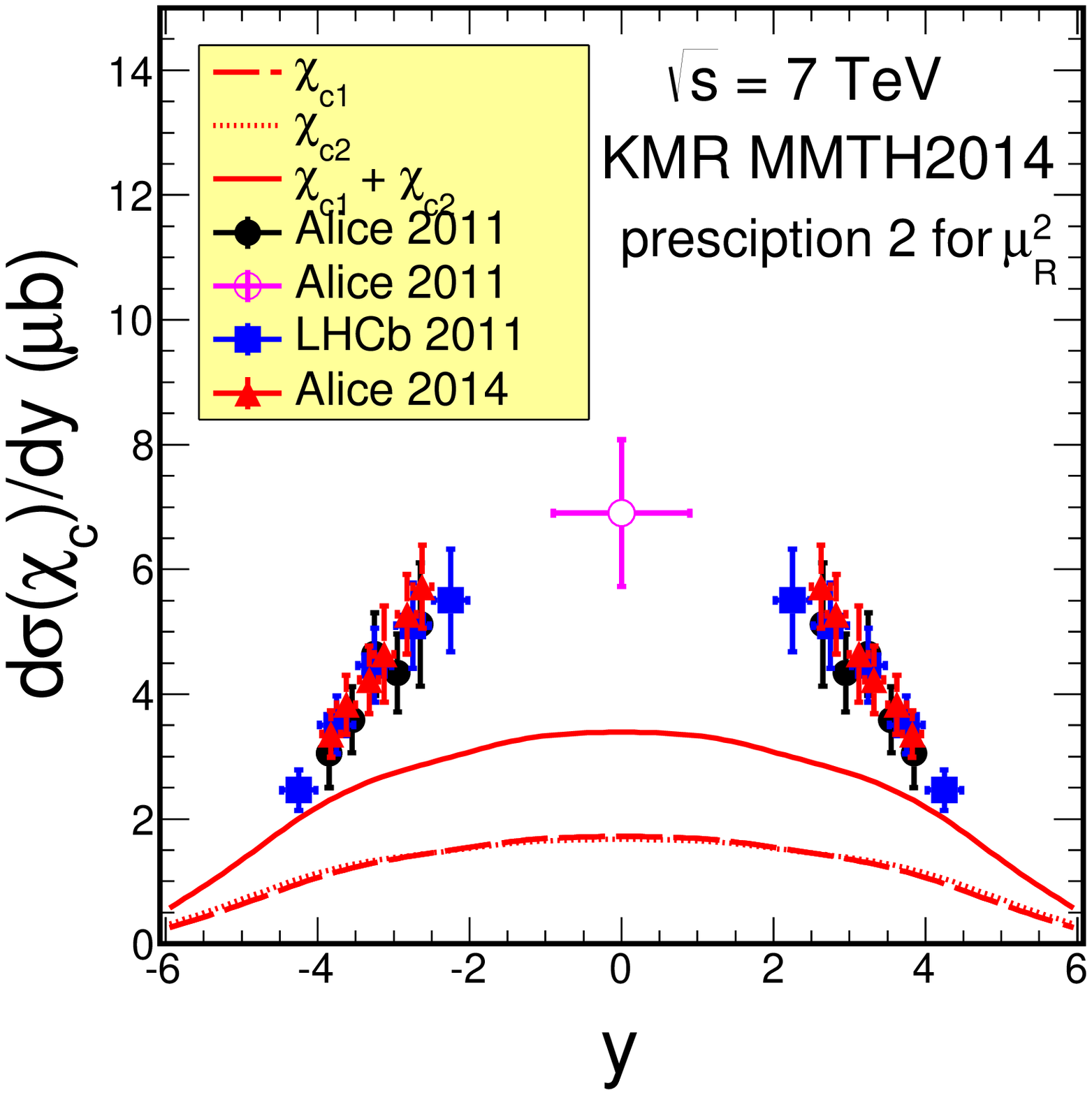}
\includegraphics[width=5.0cm]{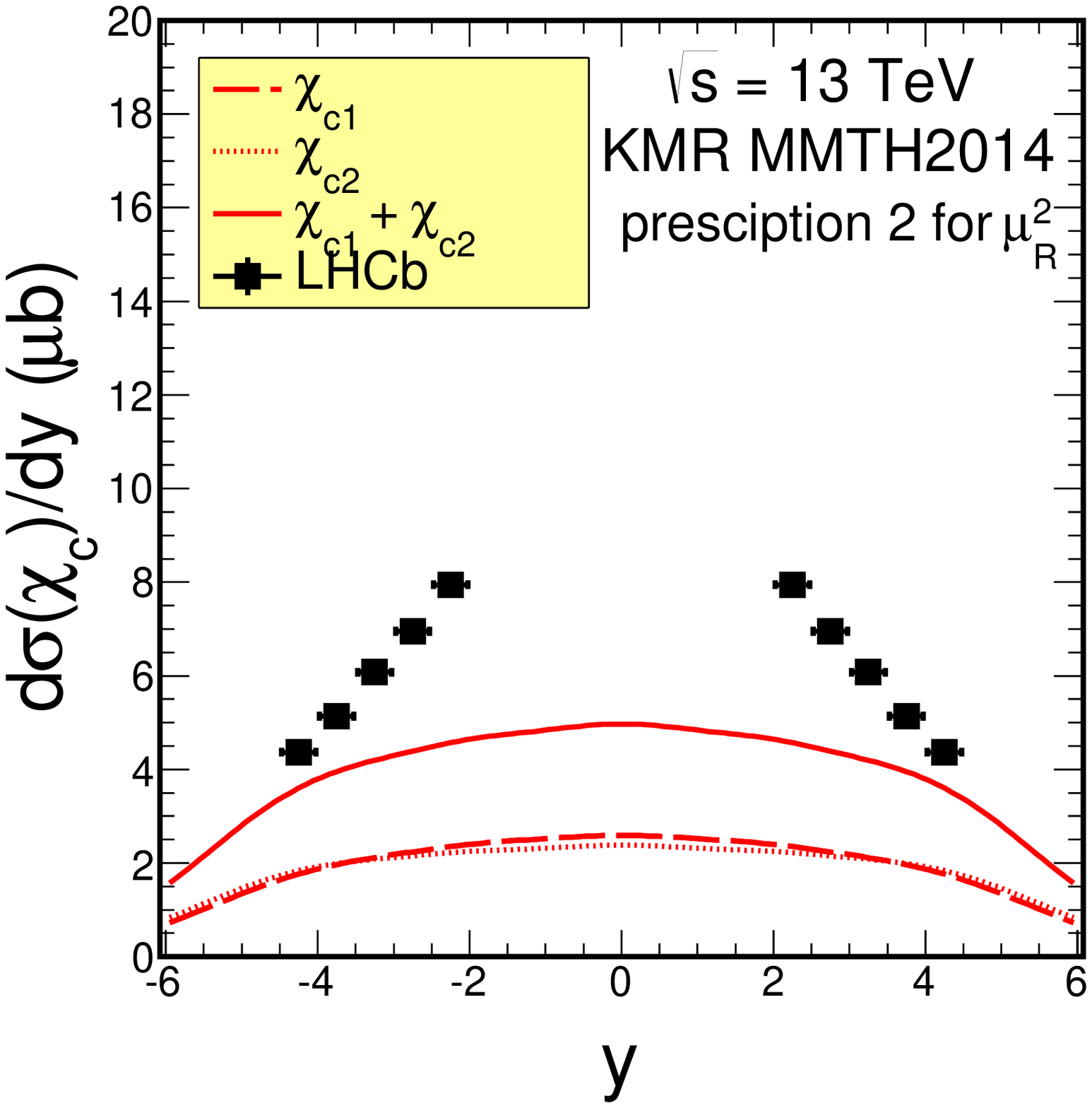}
\caption{
Rapidity distribution of $J/\psi$ mesons (from $\chi_c$ decays)
for the KMR UGDF for $\sqrt{s} =$ 2.76 GeV (left panel), 
$\sqrt{s} =$ 7 TeV (middle panel)
and $\sqrt{s} =$ 13 TeV (right panel).
The upper plots are for the scale prescription 1 and the lower plots are
for prescription 2.
}
\label{fig:dsig_dy_chic_KMR}
\end{figure}

Now we proceed to the important contribution of $J/\psi$ originating
from the feed down from the $\chi_c$ production and decay. 
The $\chi_c(0)$ meson has very small branching fraction for decay
$\chi_c(0) \to J/\psi \gamma$ (BR = 0.0127 \cite{PDG}). Therefore
in the following we shall take into account only production and
decays of $\chi_c(1)$ (BF = 0.339 \cite{PDG}) and $\chi_c(2)$ 
(BF = 0.192 \cite{PDG}). In Fig.\ref{fig:dsig_dy_chic_KMR} we show
rapidity distributions of resulting $J/\psi$ mesons for two
different prescriptions of $\alpha_s$ (compare top and bottom panels)
for three different energies $\sqrt{s}$ = 2.76, 7, 13 TeV.
This calculations were performed with the KMR UGDF.
For comparison we present also existing data of the ALICE and LHCb
collaborations. 
We show both contributions of each of the mesons and a sum
of them.
The first prescription leads to clearly too large
cross section, having in mind the other missing contributions.
This is especially clearly seen for $\sqrt{s}$ = 13 TeV, at large
rapidities. The second prescription is not in conflict with the
data, but one should remember other, not yet included, contributions
(direct one and $\psi'$ feed down).

\begin{figure}
\includegraphics[width=5.0cm]{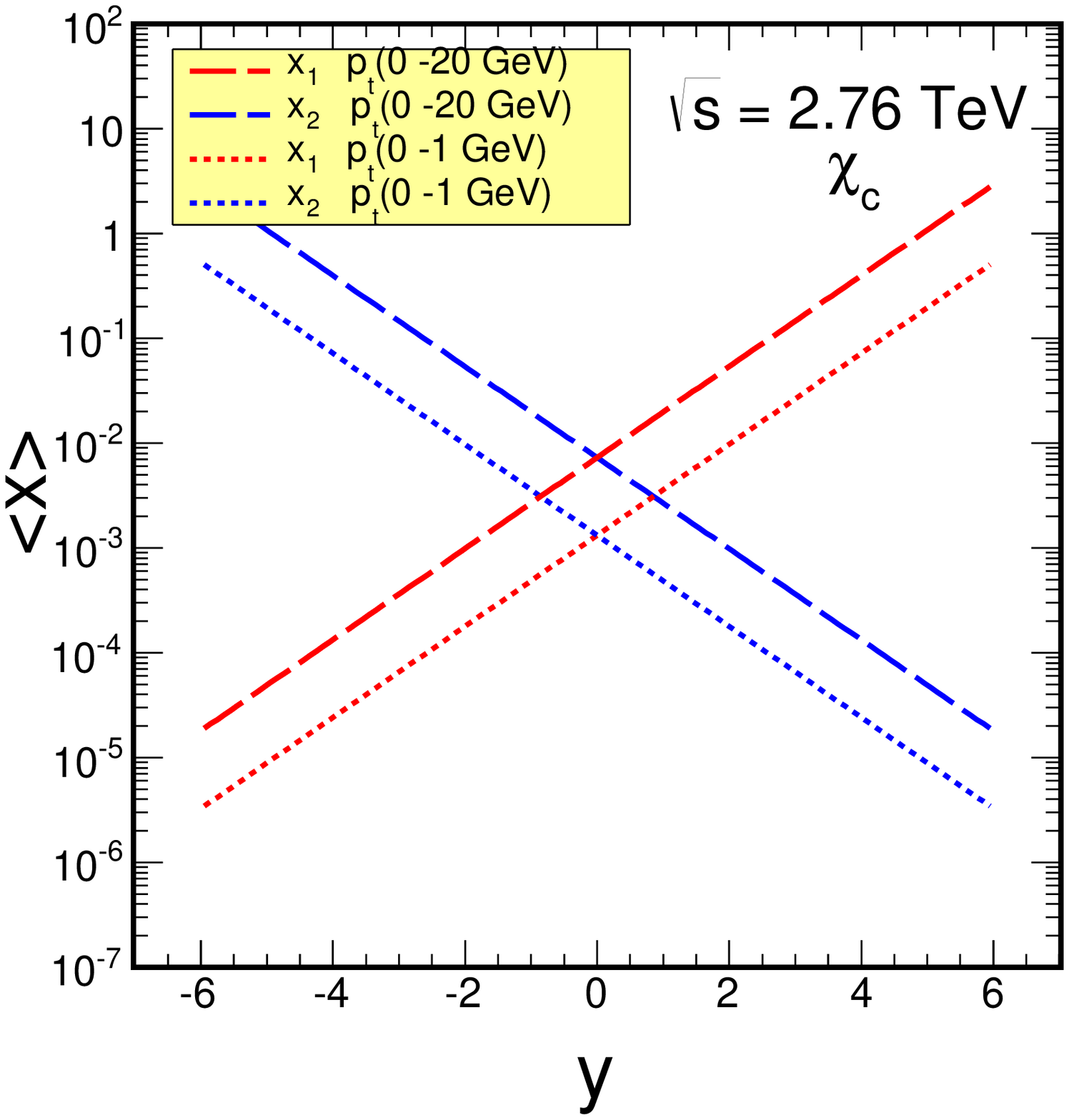}
\includegraphics[width=5.0cm]{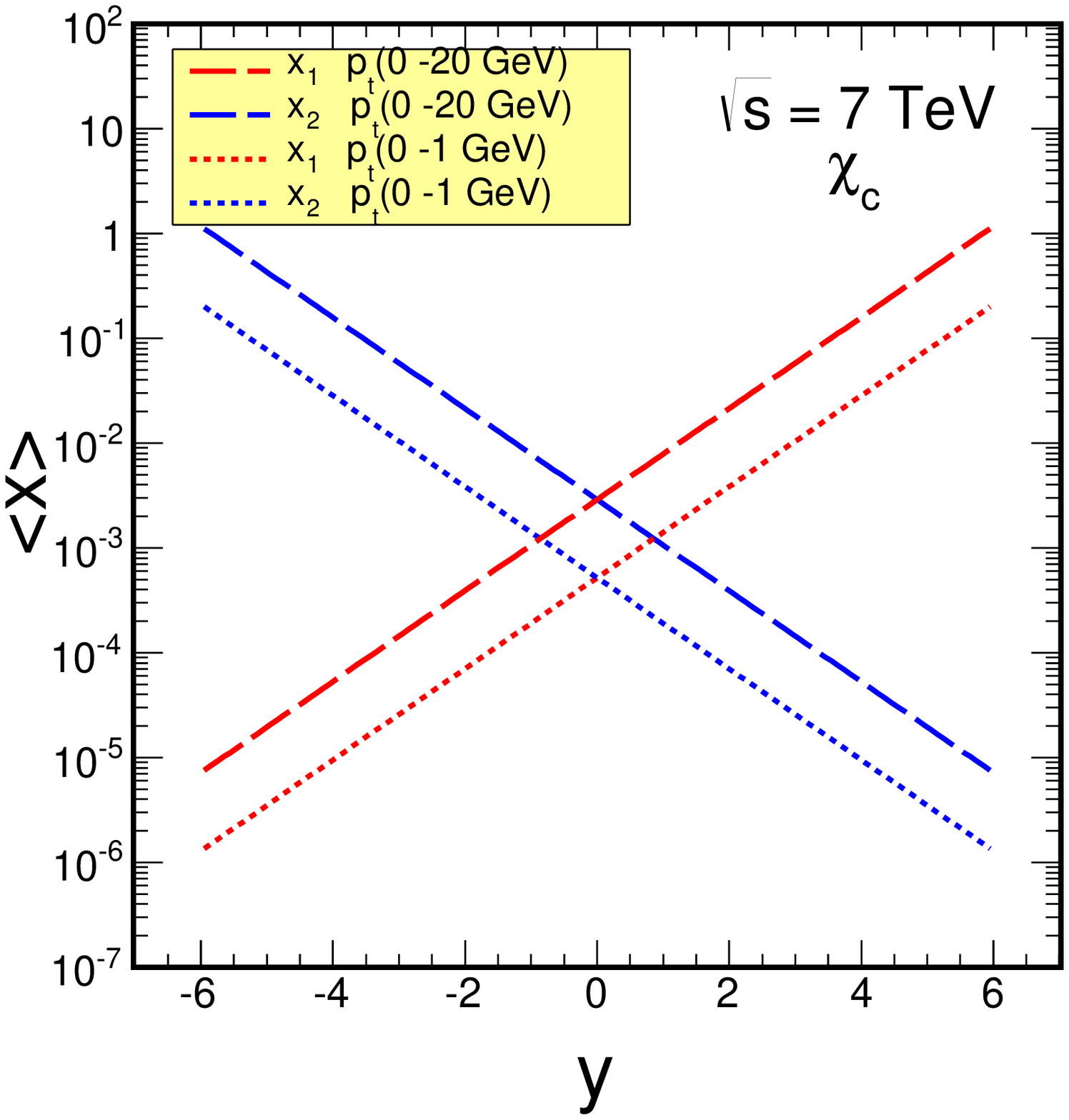}
\includegraphics[width=5.0cm]{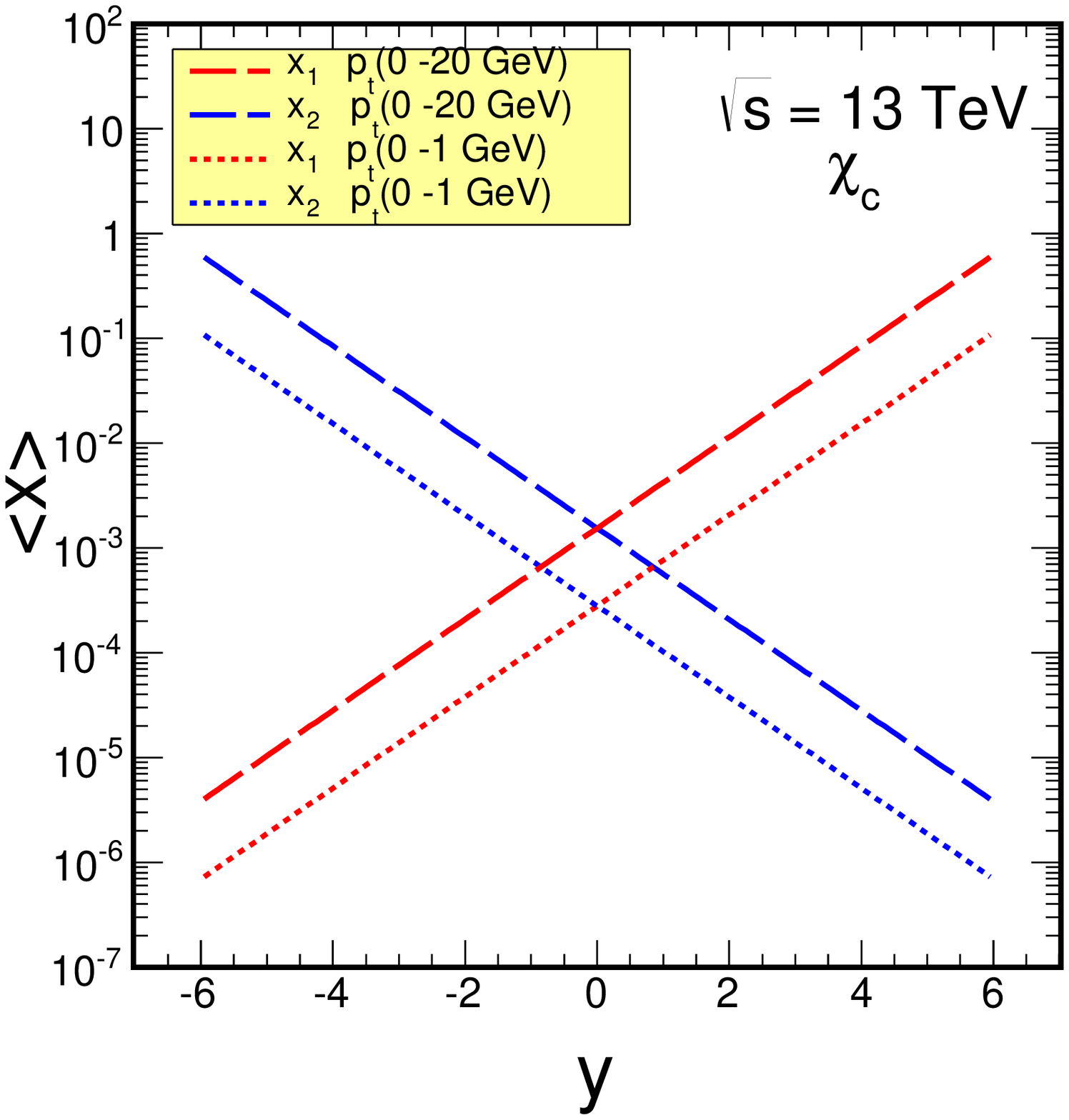}
\caption{
Average longitudinal momentum fractions $x_{1}$ and $x_{2}$ as a function of y
(rapidity of $\chi_{c}$) for  different ranges of $\chi_{c}$ transverse momenta
as specified in the figures.
}
\label{fig:x1x2_average_chic}
\end{figure}

The situation with $\chi_c$ production seems more problematic
than for the direct contribution and not yet discussed $\psi'$ feed
down. What is specific for $\chi_c$ production?
In Fig.\ref{fig:x1x2_average_chic} we show averaged values
of $x_1$ and $x_2$ being arguments of UGDFs.
Clearly in the forward LHCb rapidity region the corresponding longitudinal momentum
fractions are exteremely small. For $\sqrt{s}$ = 13 TeV they reach
the gluon longitudinal momentum fractions as small as
$x \sim$ 10$^{-6}$. This makes the forward production of $\chi_{c}$ very special
in the context of searching for saturation/nonlinear effects.
We show results when the averaging is performed in different regions of
$\chi_{c}$ transverse momenta.

\begin{figure}
\includegraphics[width=5.0cm]{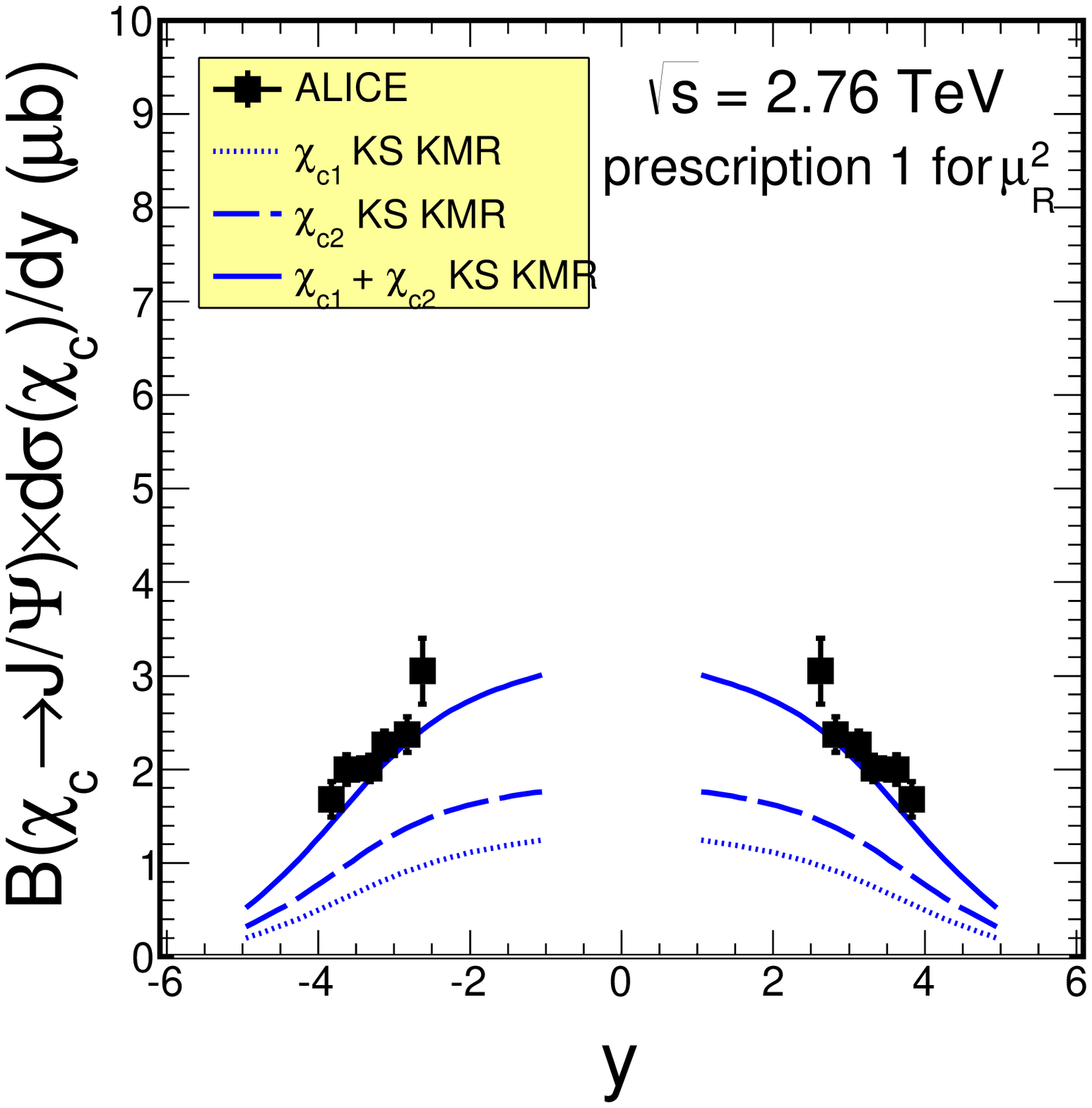}
\includegraphics[width=5.0cm]{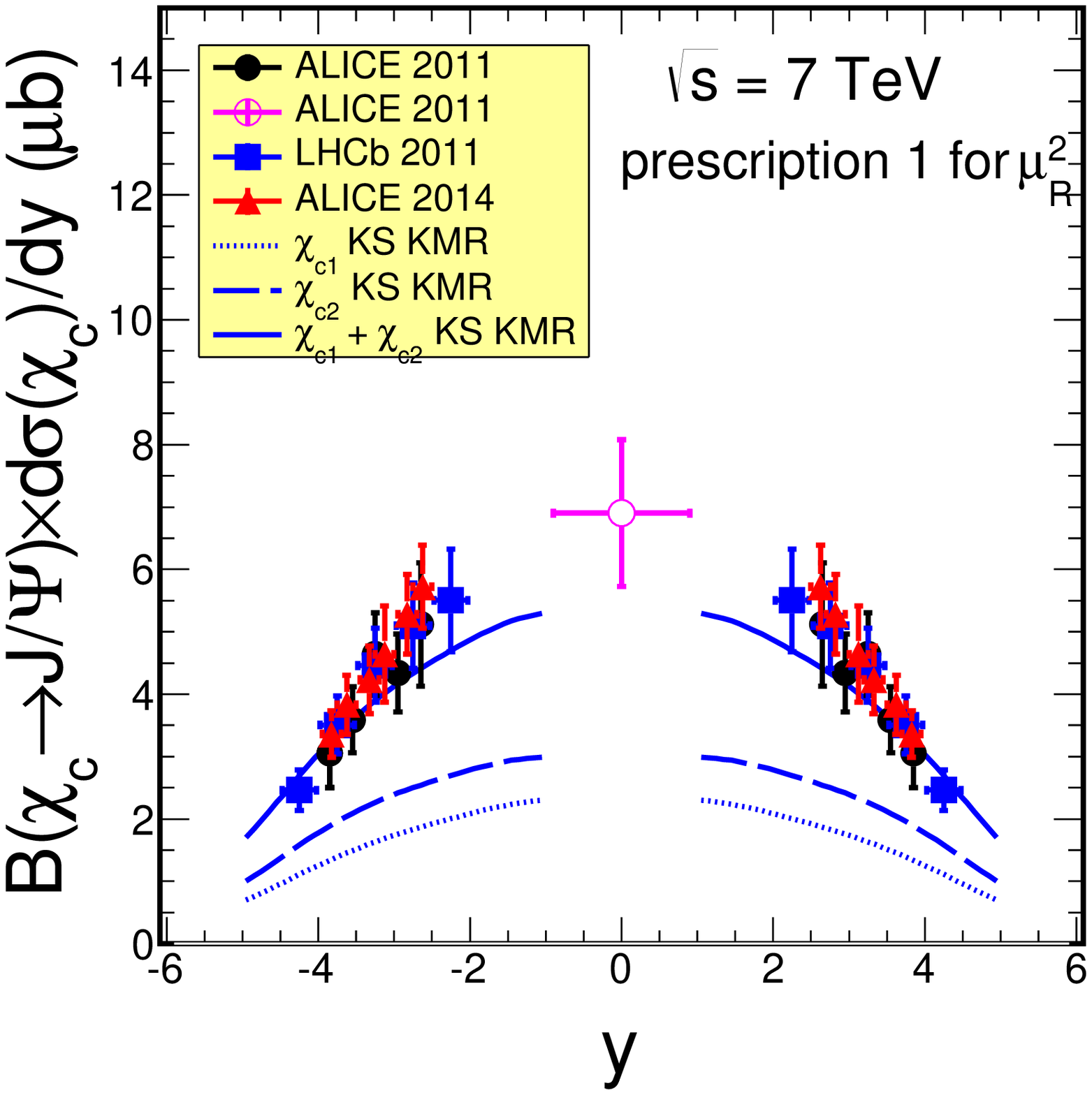}
\includegraphics[width=5.0cm]{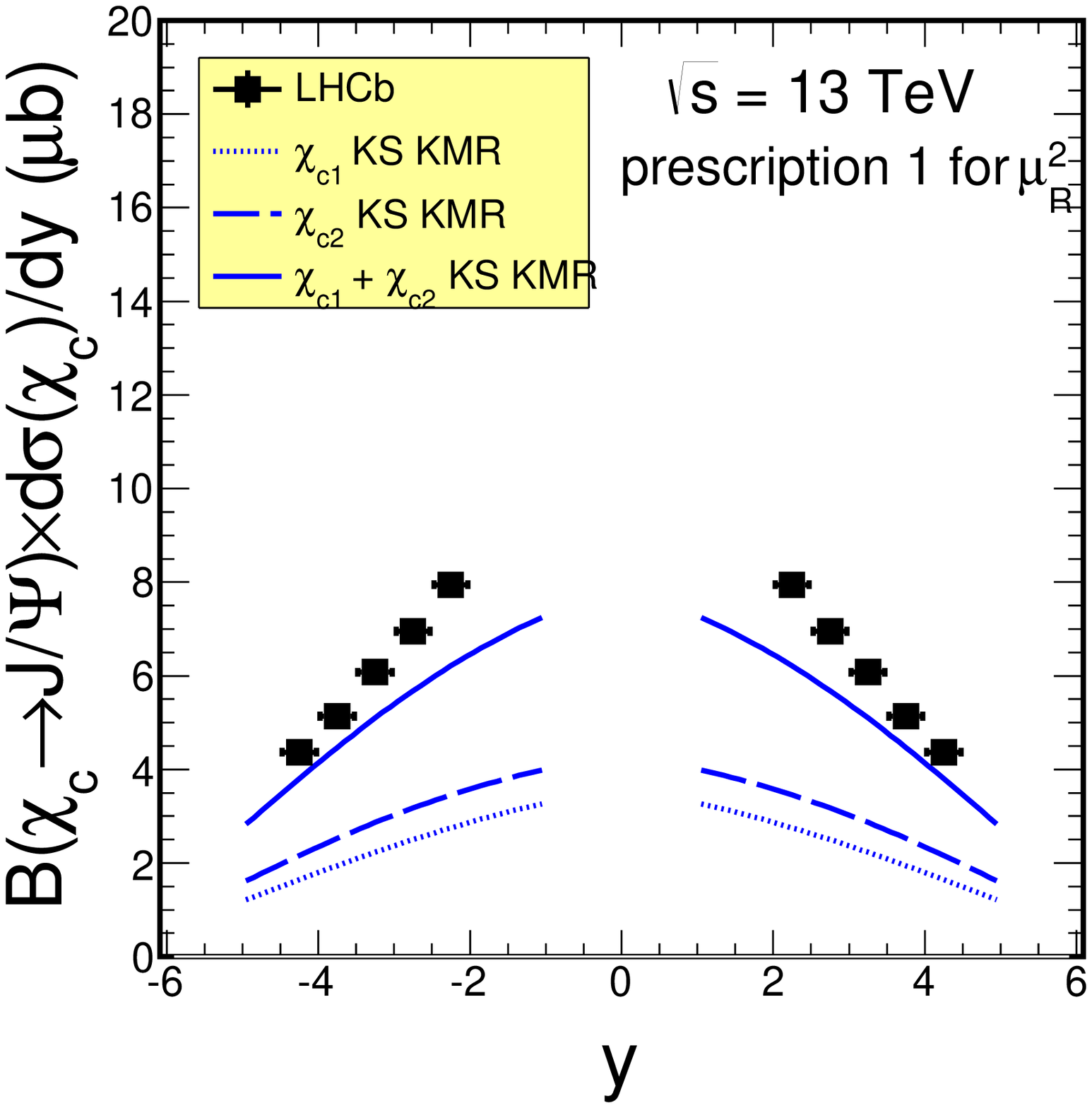} \\
\includegraphics[width=5.0cm]{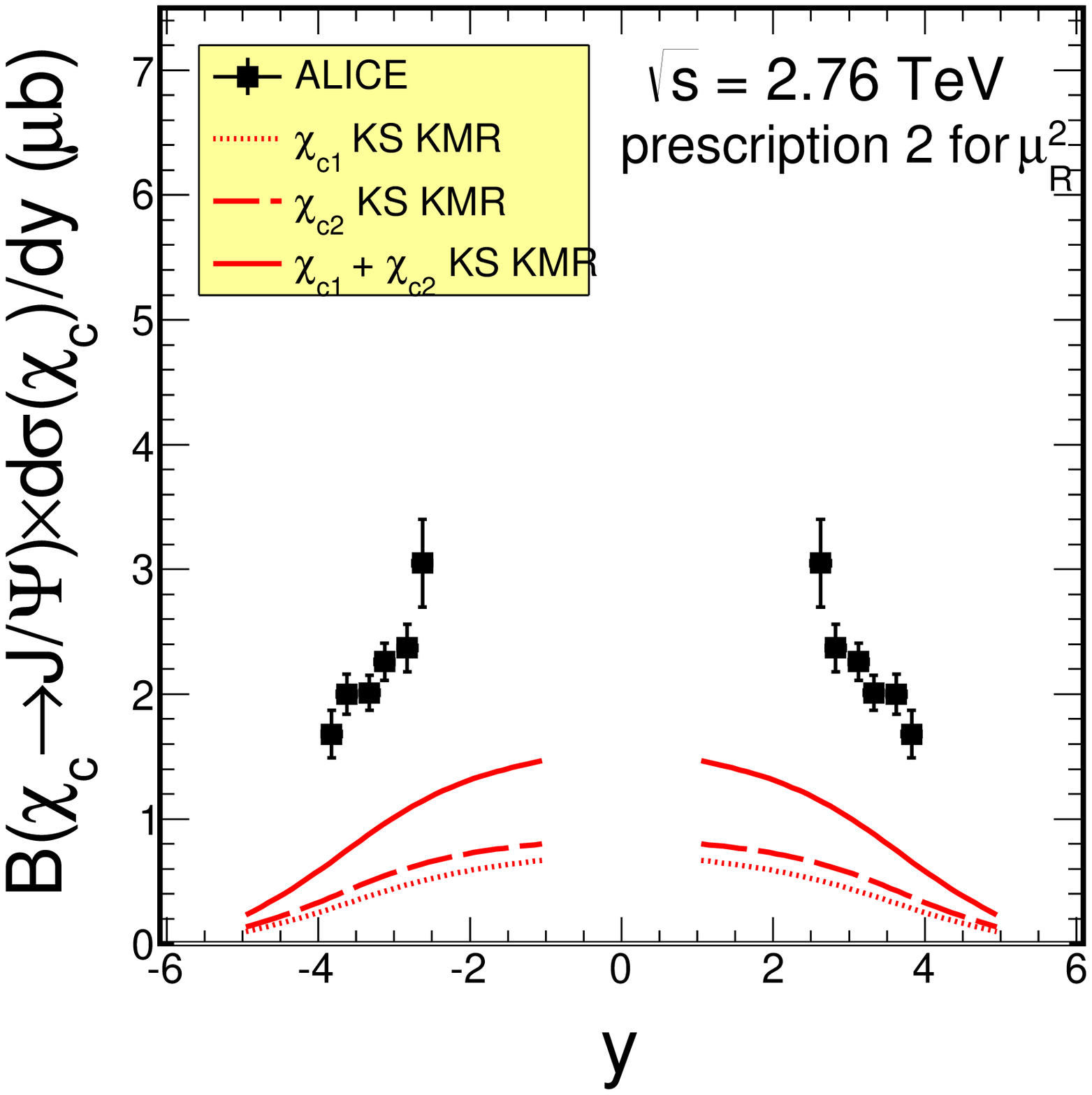}
\includegraphics[width=5.0cm]{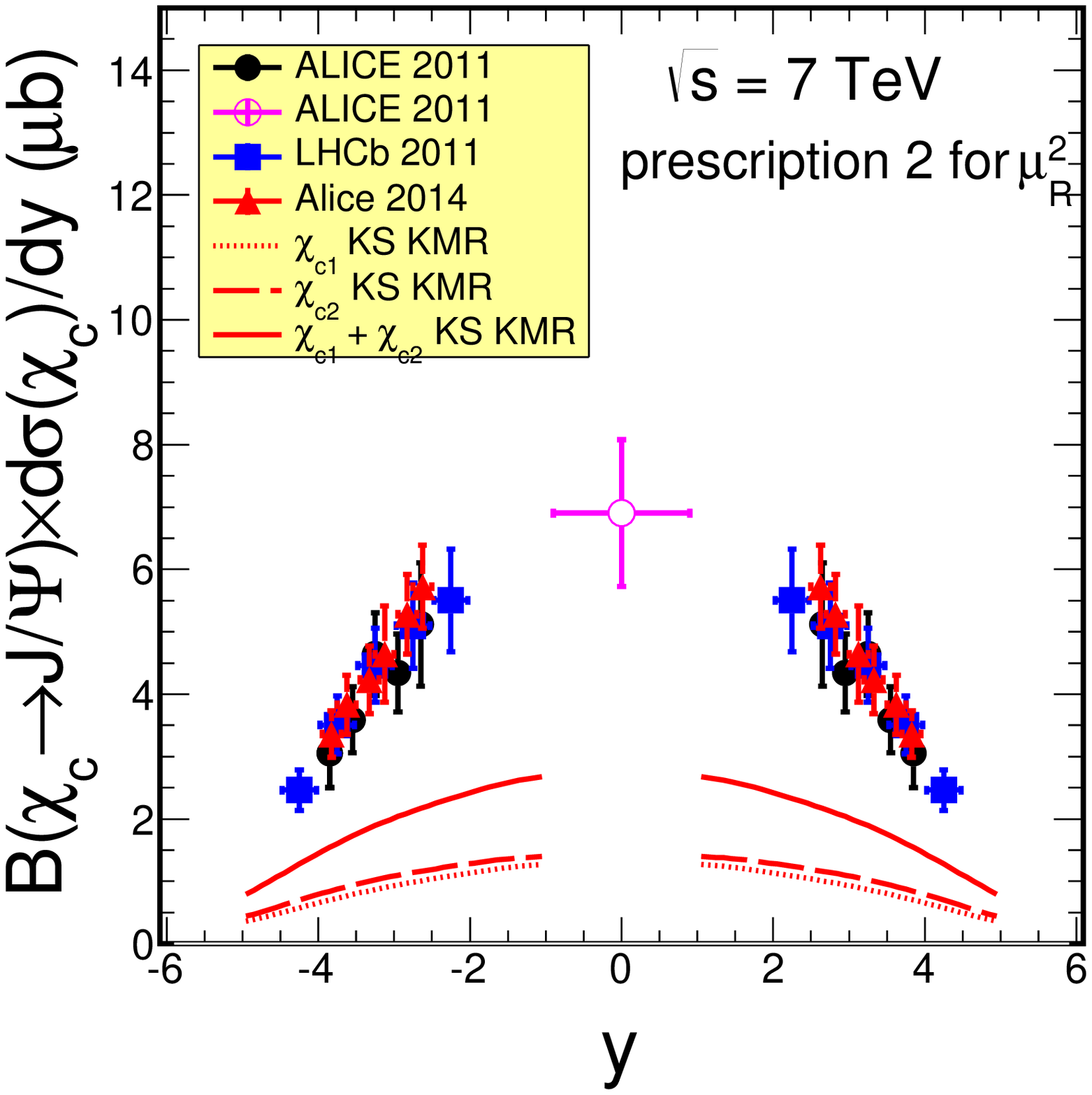}
\includegraphics[width=5.0cm]{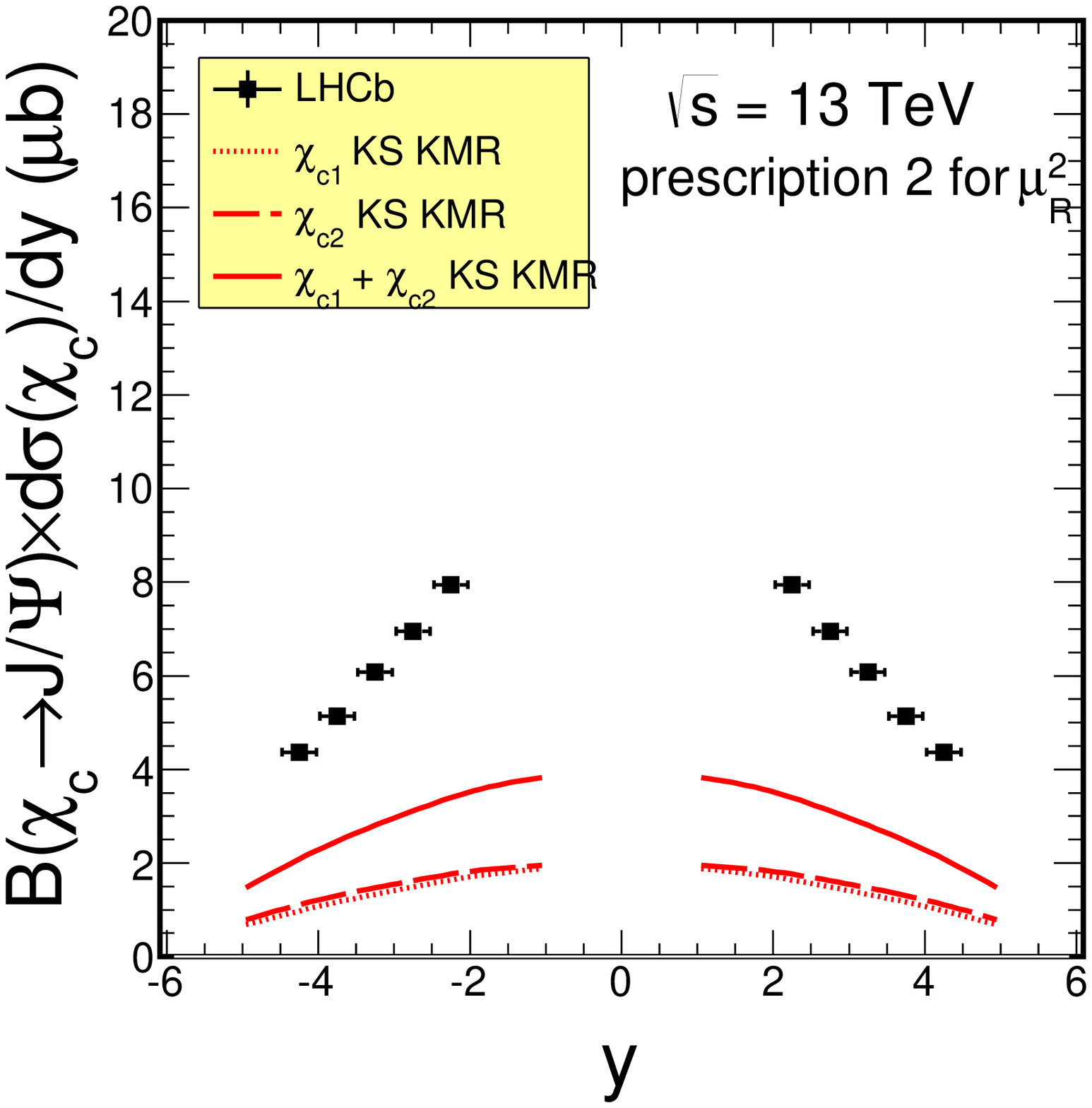}
\caption{
Rapidity distribution of $J/\psi$ mesons (from $\chi_c$ decays)
for the mixed UGDFs for $\sqrt{s} =$ 2.76 GeV (left panel), 
$\sqrt{s} =$ 7 TeV (middle panel)
and $\sqrt{s} =$ 13 TeV (right panel).
The upper plots are for the scale prescription 1 and the lower plots are
for prescription 2.
}
\label{fig:dsig_dy_chic_mixed}
\end{figure}

The very small values of longitudinal momentum fractions
relevant for $\chi_c$ production in the forward directions fully justify
the use of the ``mixed'' UGDFs, discussed already in the context of 
direct production. In Fig.\ref{fig:dsig_dy_chic_mixed} we show
corresponding rapidity distributions. The results obtained for
the ``mixed'' distributions are quite different than those obtained
solely with the KMR UGDFs, especially for $\sqrt{s}$ = 13 TeV.
Is it a sign of the onset of saturation? This should be clarified in future
by dedicated measurements of $\chi_c$ mesons for different rapidities.
This process seems to be very promissing in this context.

\subsection{$\chi_c$ production}

\begin{figure}
\includegraphics[width=5.0cm]{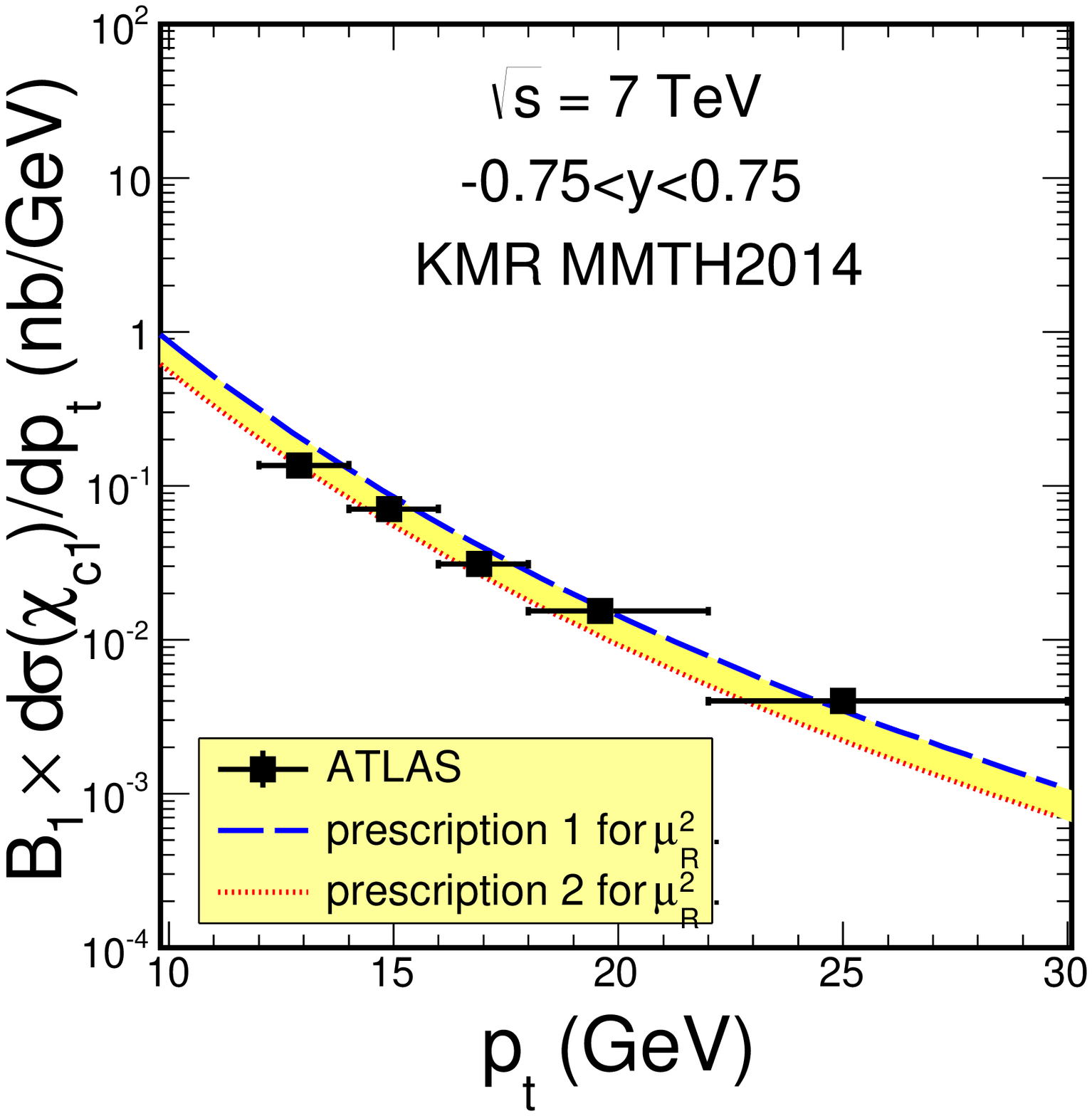}
\includegraphics[width=5.0cm]{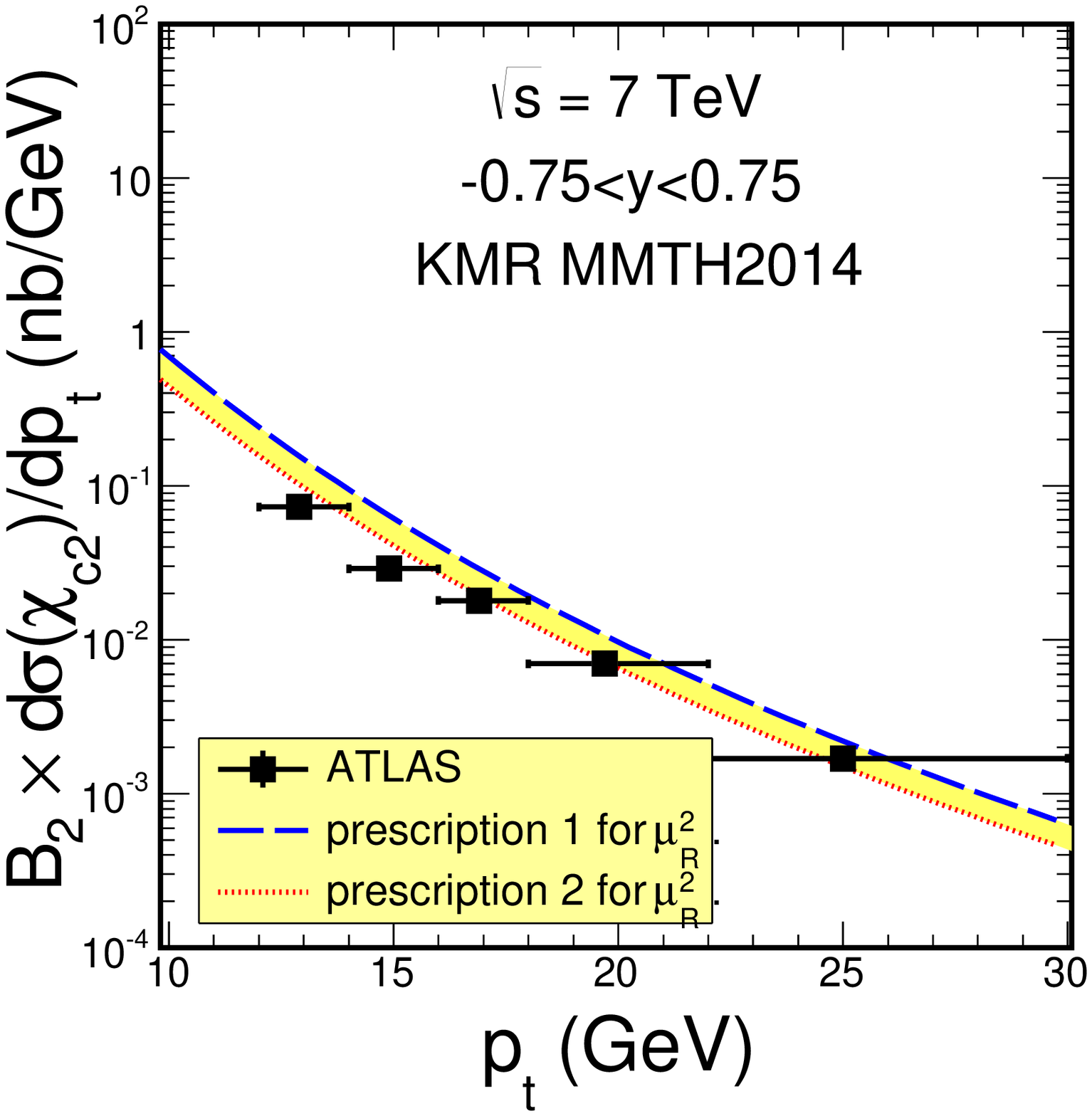}
\includegraphics[width=5.0cm]{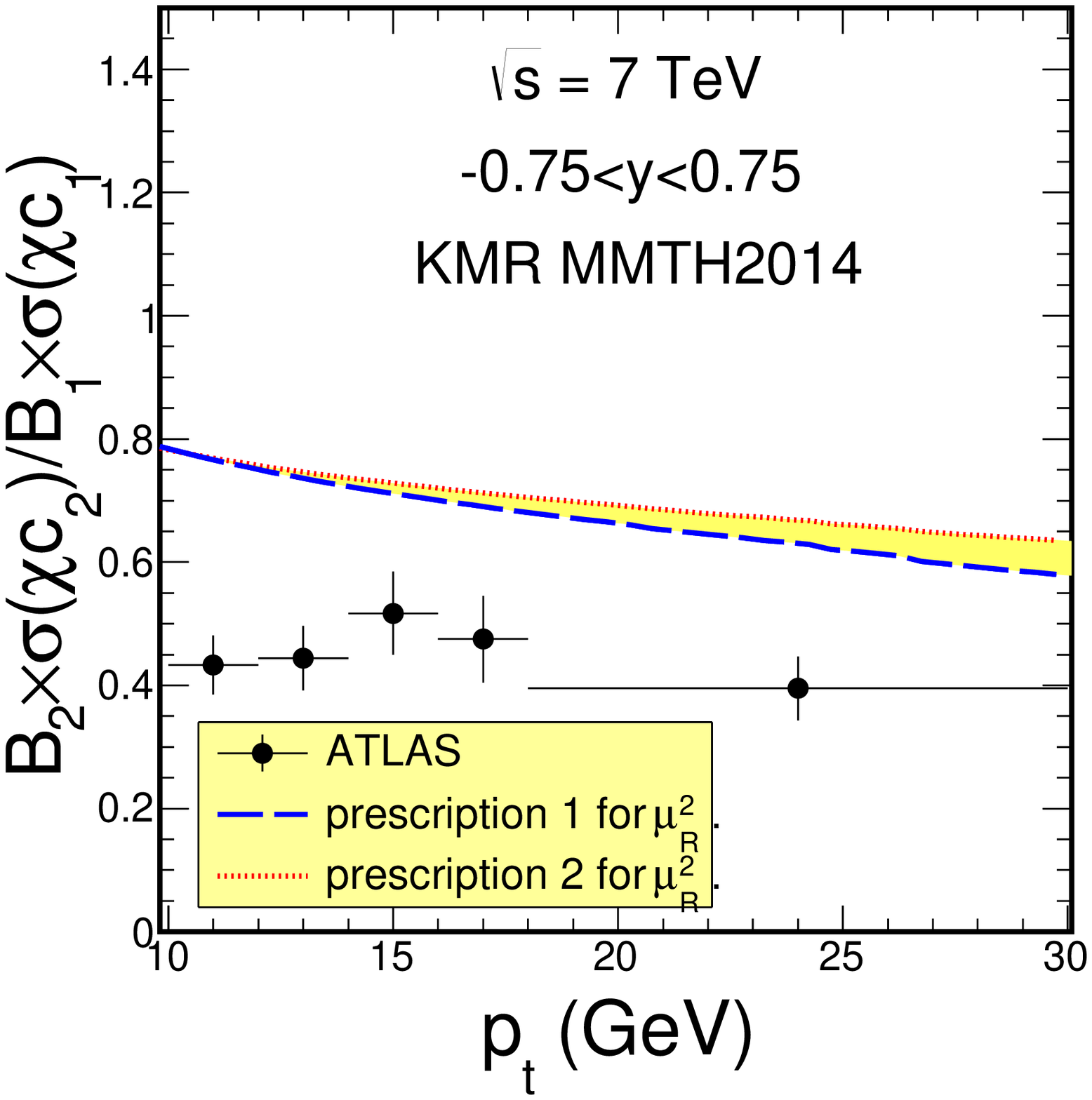}\\
\includegraphics[width=5.0cm]{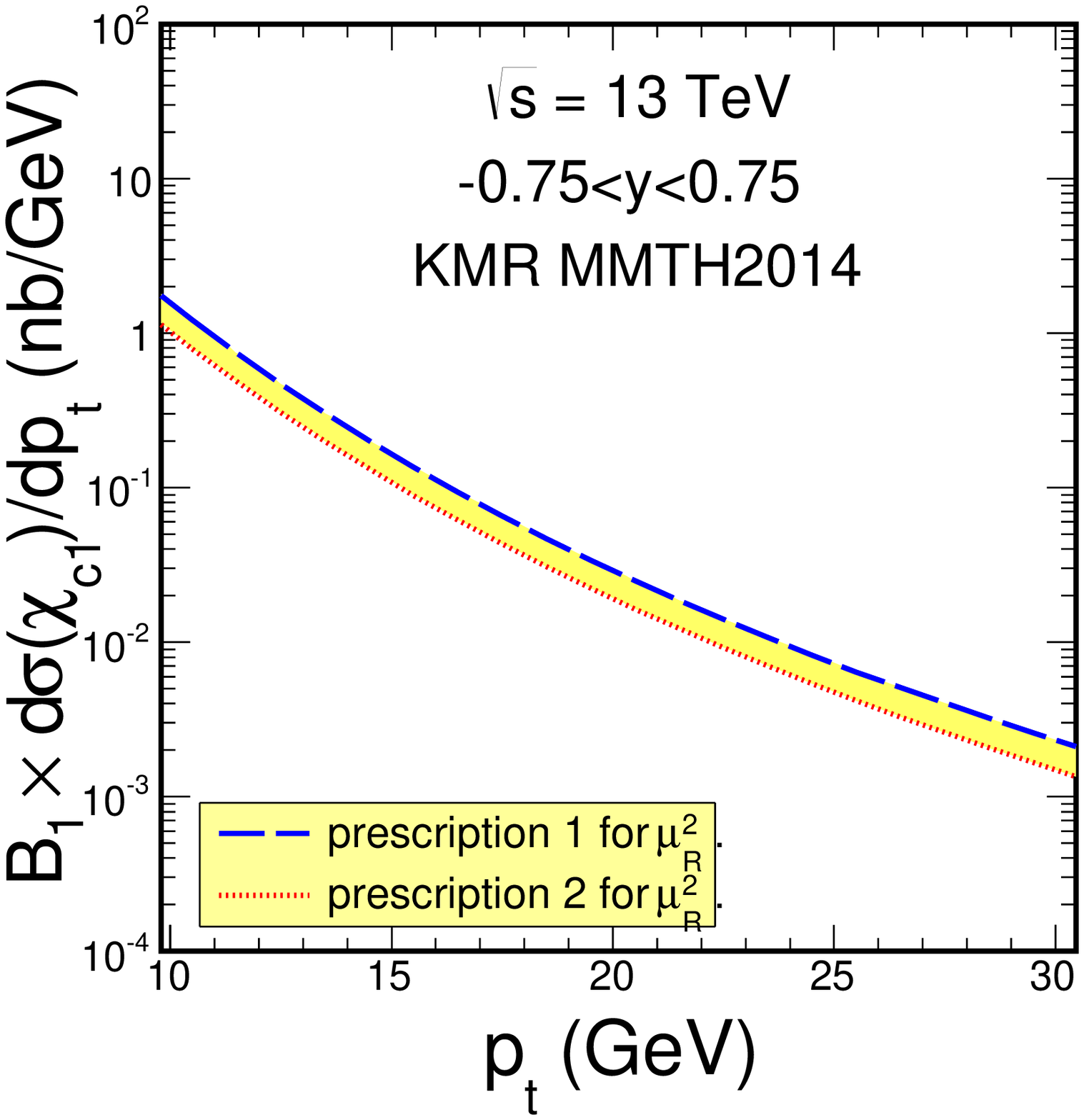}
\includegraphics[width=5.0cm]{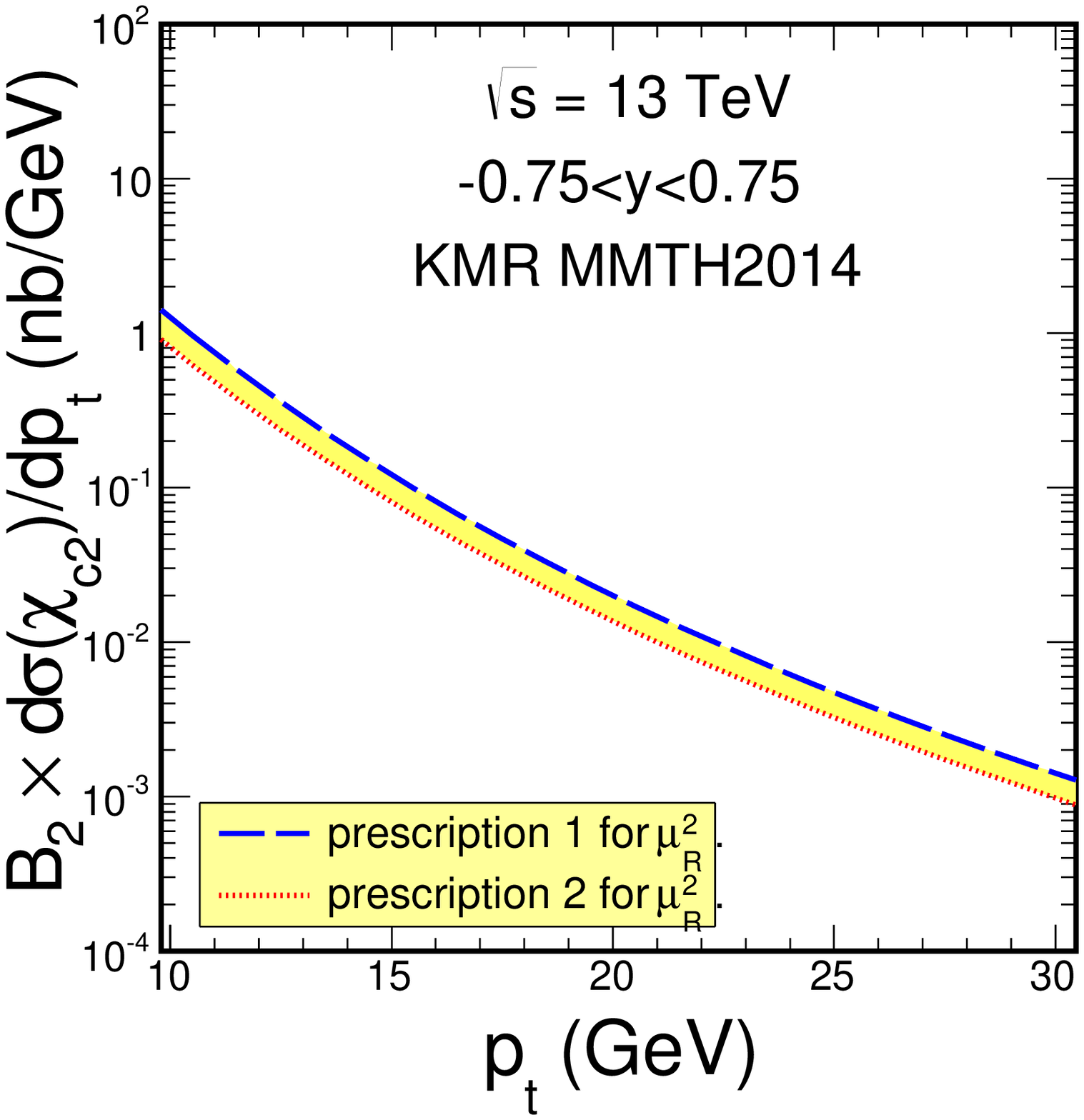}
\includegraphics[width=5.0cm]{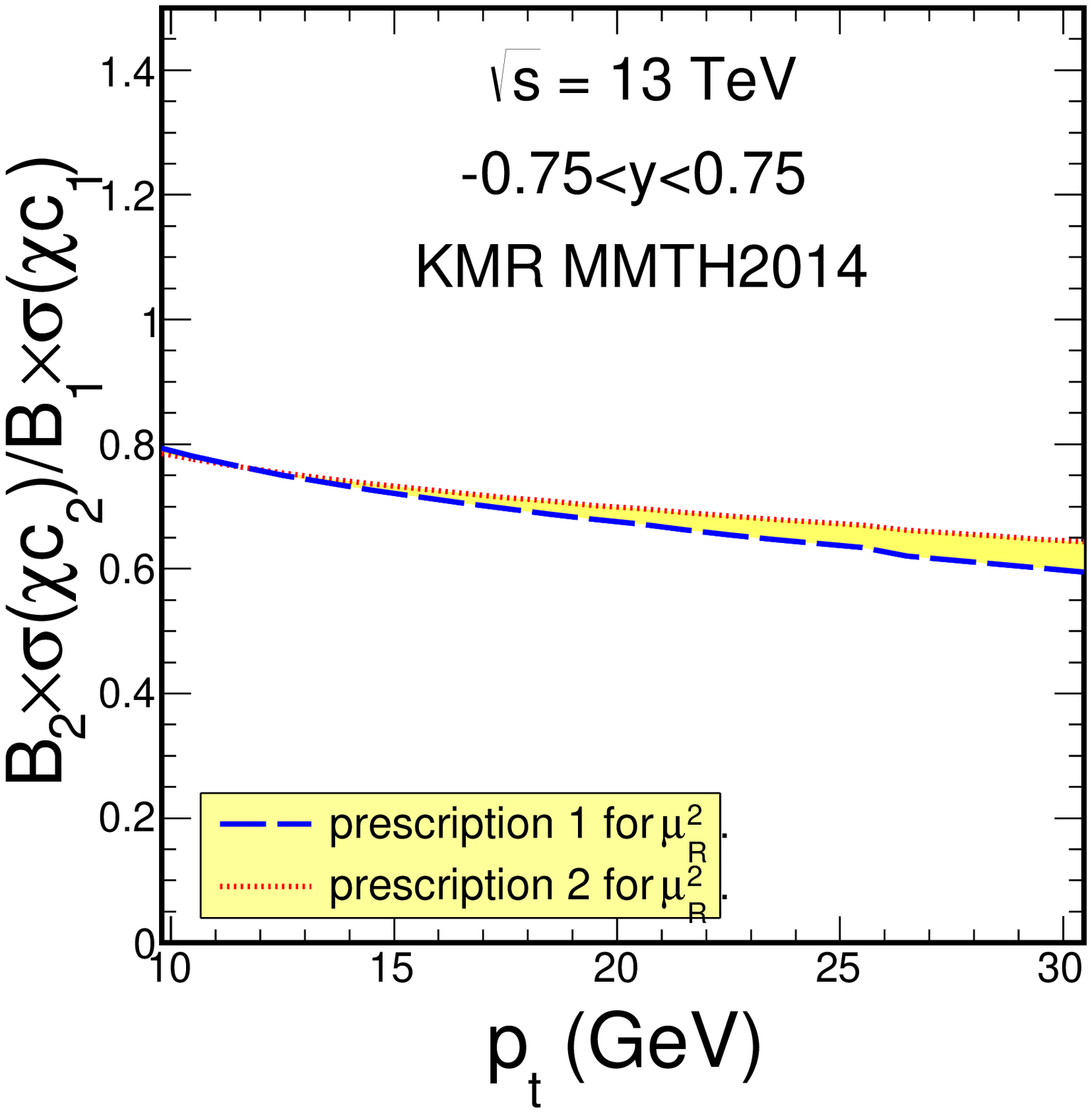}
\caption{
Transverse momentum distributions of $\chi_c(1)$ (left panels)
and $\chi_c(2)$ (middle panels). The right panels show
the ratio of $\chi_c(2)$ to $\chi_c(1)$.
The upper plots are for $\sqrt{s} =$ 7 TeV and the lower plots
are for $\sqrt{s} =$ 13 TeV.
The experimental ATLAS \cite{ATLAS_2014} data are shown for comparison.
}
\label{fig_dsig_dpt_chic}
\end{figure}

So far $\chi_c$ mesons were measured only at $\sqrt{s}$ = 7 TeV, 
at midrapidities and rather large transverse momenta. 
Then the corresponding longitudinal momentum fractions are not so
small. 
In Fig.\ref{fig_dsig_dpt_chic} we show our results for both $\sqrt{s}$ = 7 TeV, 
together with ATLAS experimental data 
\cite{ATLAS_2014}, and our predictions for $\sqrt{s}$ = 13 TeV.
We get reasonable, but not ideal, description of the experimental
transverse momentum distributions for $\chi_c(1)$ (left panel) 
and $\chi_c(2)$ (middle panel). We slightly overestimate the data
for $\chi_c(2)$ especially for smaller values of transverse momenta.
For completeness we show the ratio $\chi_c(2)/\chi_c(1)$.
In principle, we could try to treat parameters of $\chi_c(1)$ and
$\chi_c(2)$ independently and better fit them to the ATLAS data, 
but we leave it for future when next-to-leading order corrections will
be included. Summarizing this short subsection, we have shown that our
parameters for $\chi_c$ are reasonable. They are to some extend
effective as only leading-order $k_t$-factorization is done here.
How it changes at next-to-leading order clearly goes beyond the scope of
the present analysis.

\subsection{All contributions for $J/\psi$ production}

Having reviewed all components separately we are ready to include 
all of them together. 
In the following we will adopt always prescription 2 (lower limits above)
for $\alpha_s$ as an example.

\begin{figure}
\includegraphics[width=5.0cm]{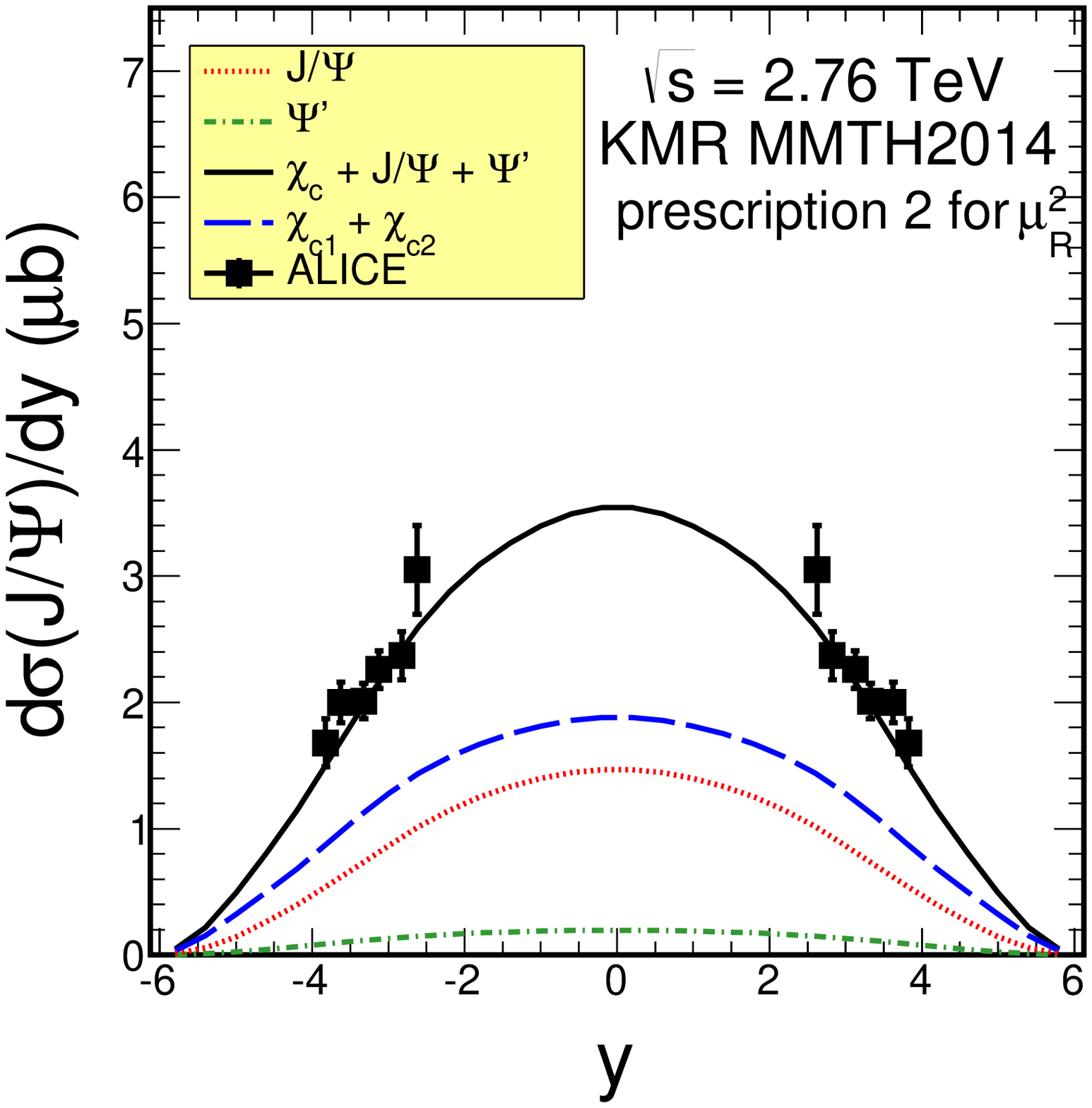}
\includegraphics[width=5.0cm]{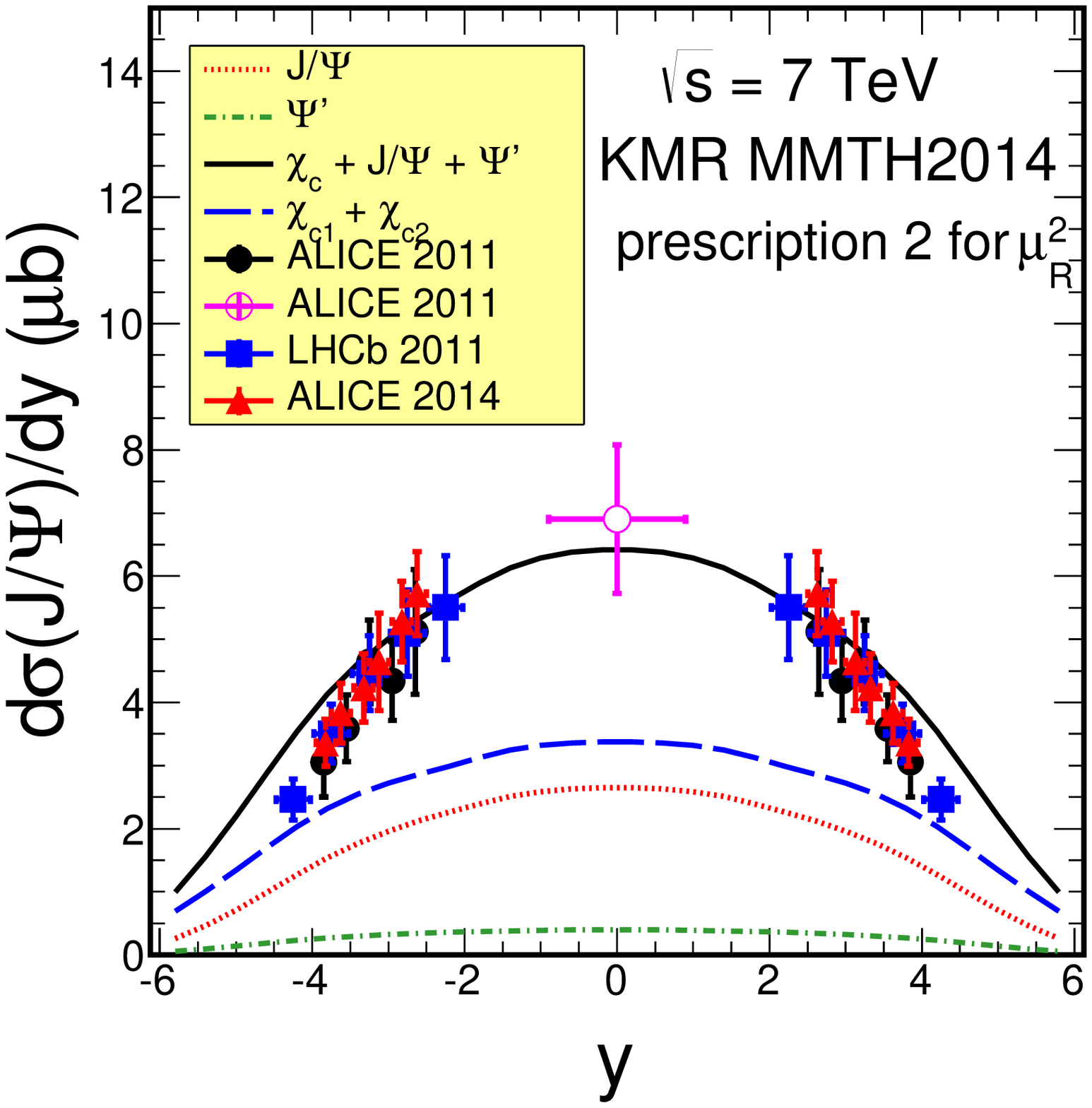}
\includegraphics[width=5.0cm]{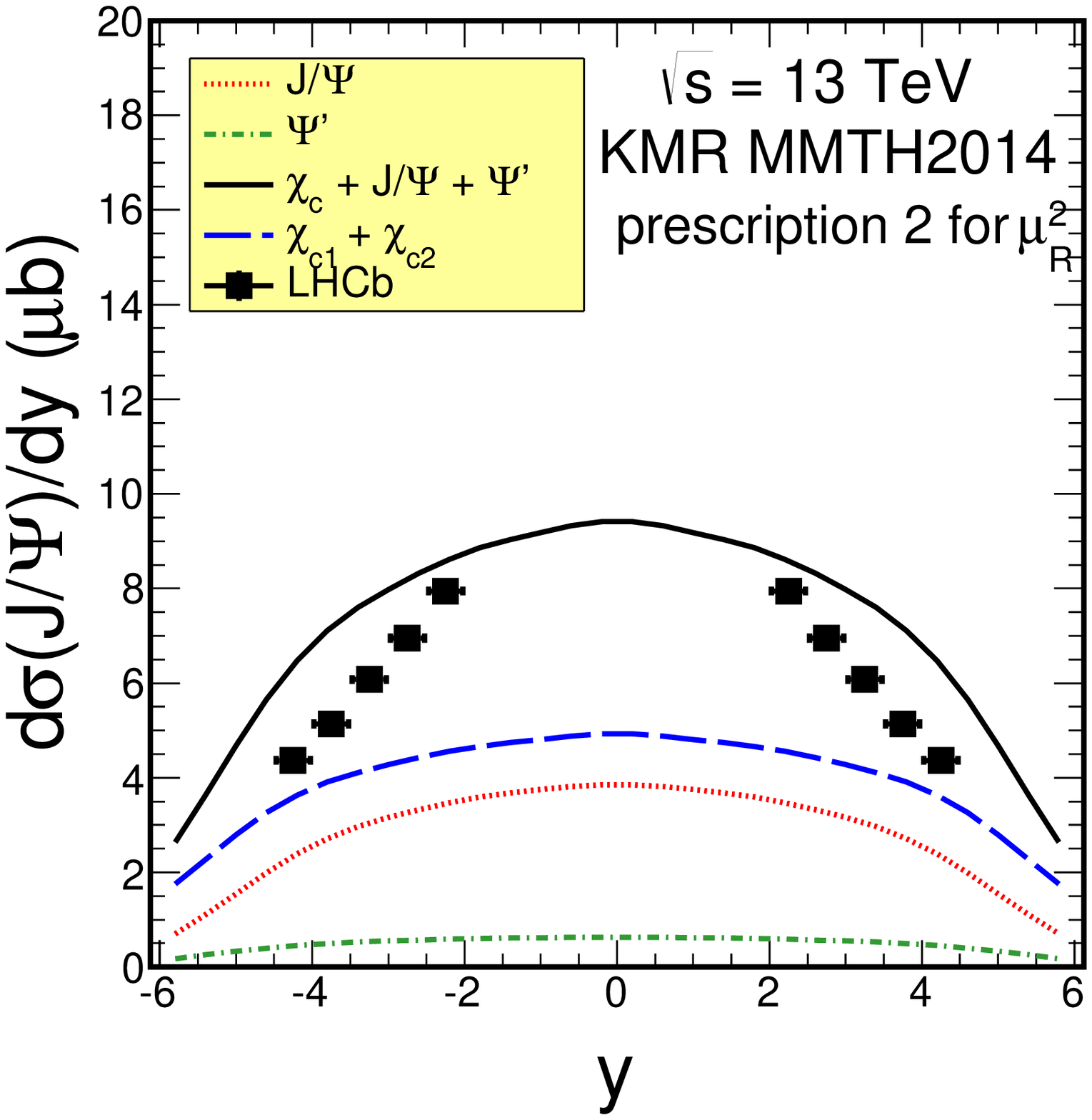}
\caption{
Rapidity distribution of $J/\psi$ mesons for all considered mechanisms
for three different energies. In this calculation for all
cases the KMR UGDF and prescription 2 were used.
}

\label{fig:dsig_dy_jpsi_all_KMR}
\end{figure}

In Fig.\ref{fig:dsig_dy_jpsi_all_KMR} we show corresponding results
for the KMR UGDF. While we get good description of the experimental
distribution for $\sqrt{s}$ = 2.76 TeV, slightly worse for 
$\sqrt{s}$ = 7 TeV, there is clear disagreement for $\sqrt{s}$ = 13 TeV.
The disagreement is larger for larger rapidities (smaller longitudinal
momentum fractions). This may be related to onset of saturation
in this region of phase space and is worth of further study.

\begin{figure}
\includegraphics[width=5.0cm]{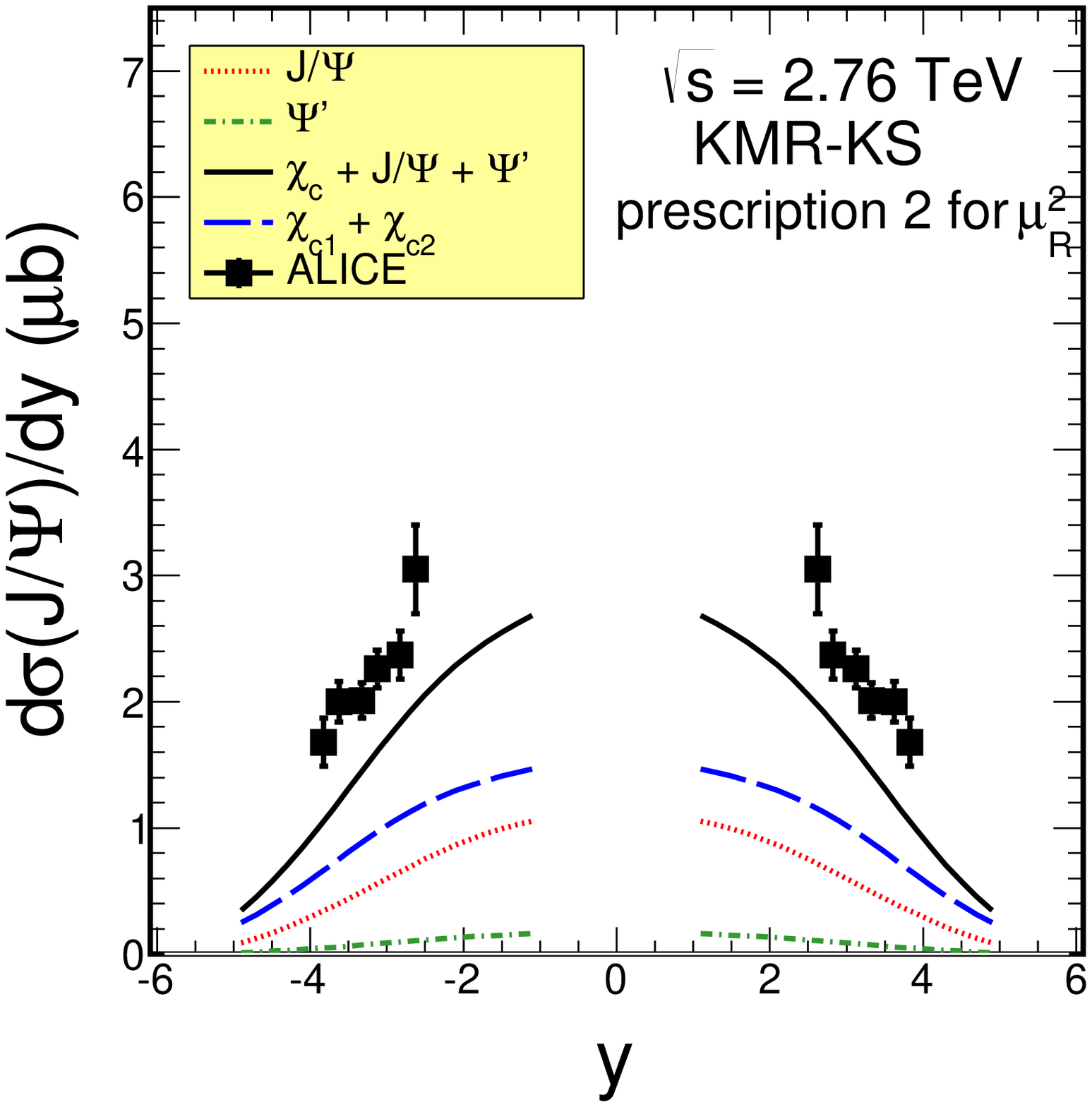}
\includegraphics[width=5.0cm]{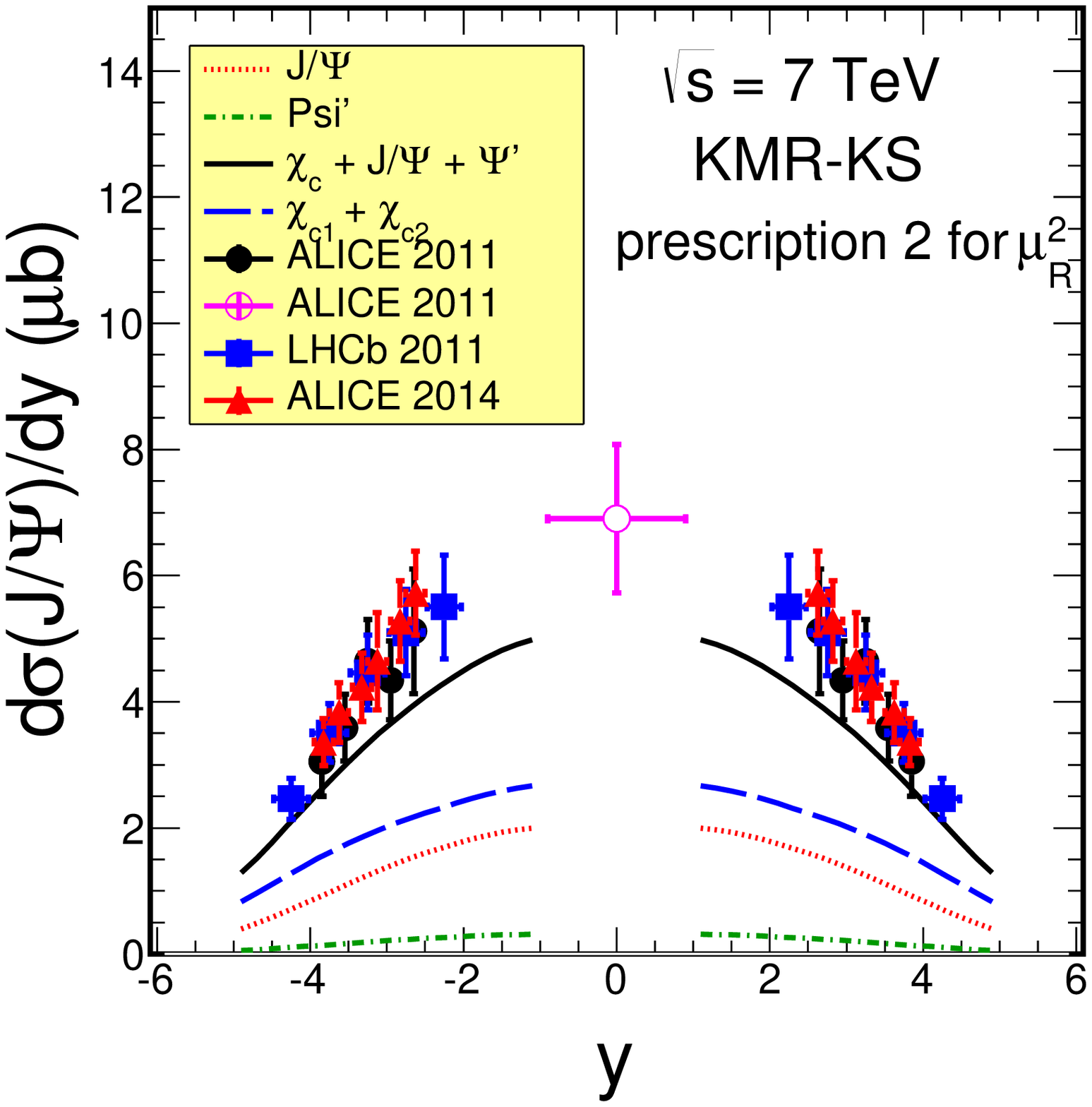}
\includegraphics[width=5.0cm]{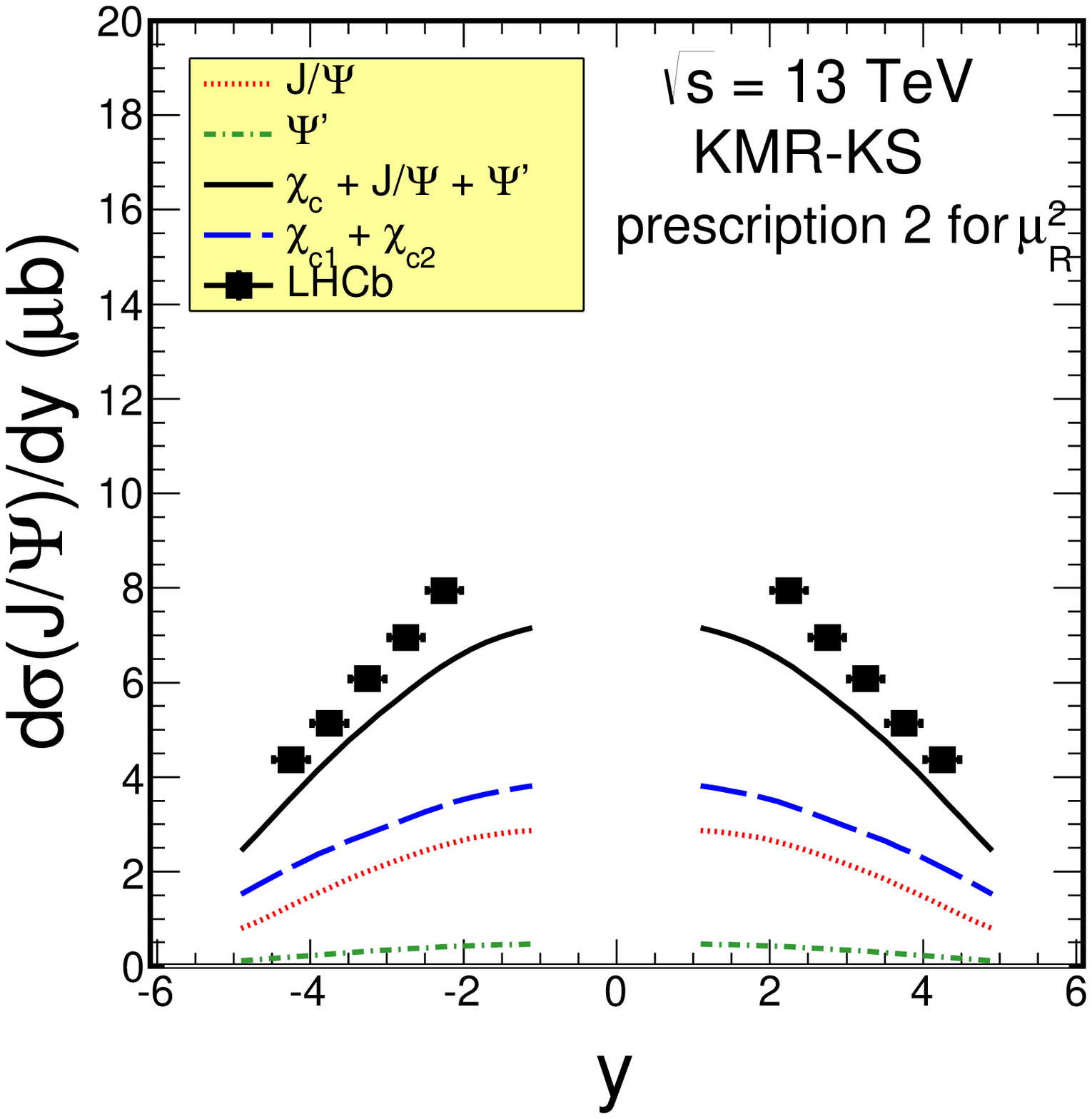}
\caption{Rapidity distribution of $J/\psi$ mesons for all considered mechanisms
for three different energies. In this calculation for all
cases ($J/\psi$, $\psi'$ and $\chi_{c}$) the mixed UGDFs scenario
and prescription 2 were used.}
\label{fig:dsig_dy_jpsi_all_mixed}
\end{figure}

In Fig.\ref{fig:dsig_dy_jpsi_all_mixed} we show similar results for
the ``mixed'' scenario. We get too much damping of the cross section,
especially for largest $\sqrt{s}$. This may signal also presence
of other, nonincluded, mechanisms or may signal that the KS saturation 
effects are too strong. They may also appear too early in $x$.

\begin{figure}
\includegraphics[width=5.0cm]{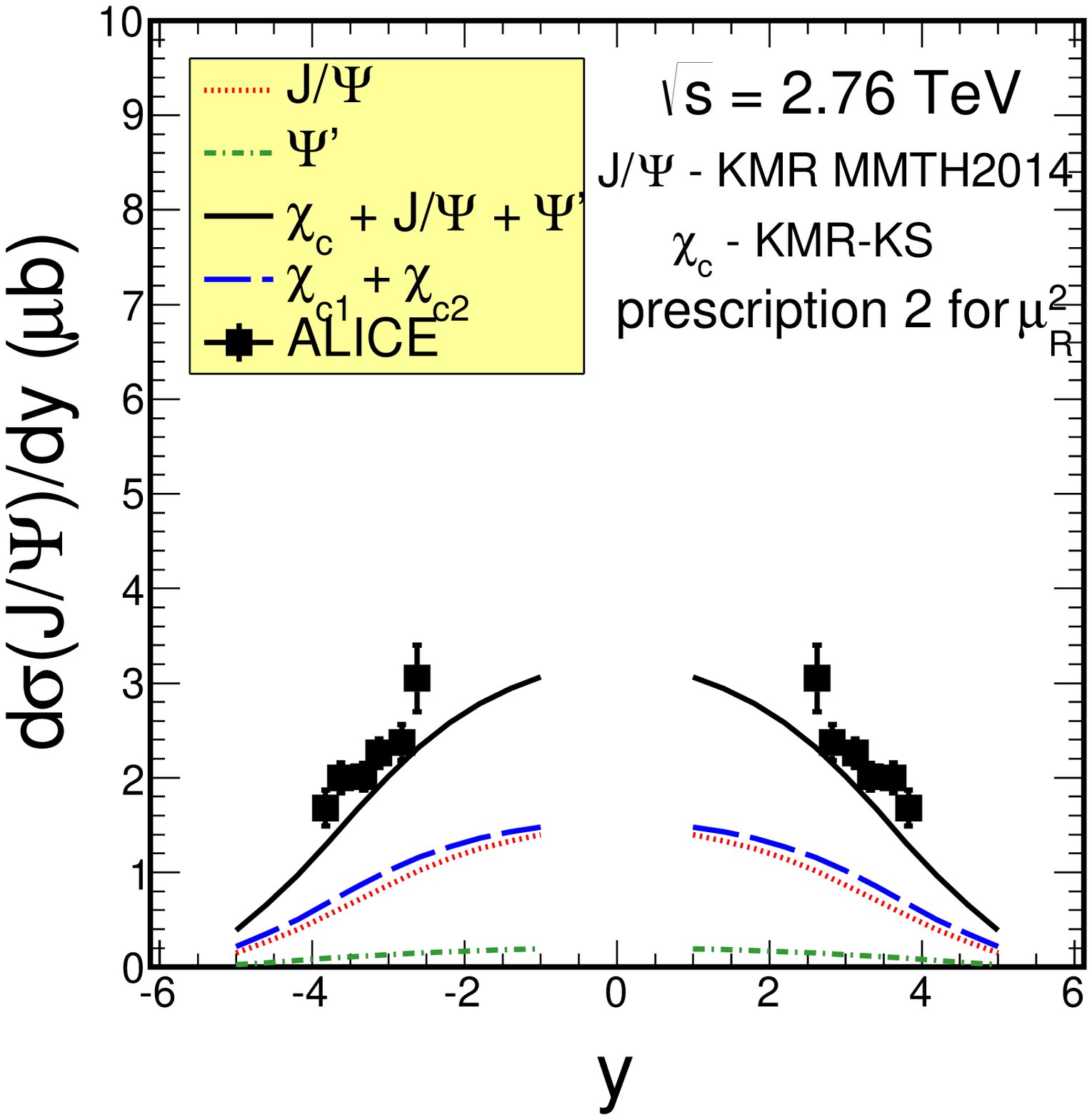}
\includegraphics[width=5.0cm]{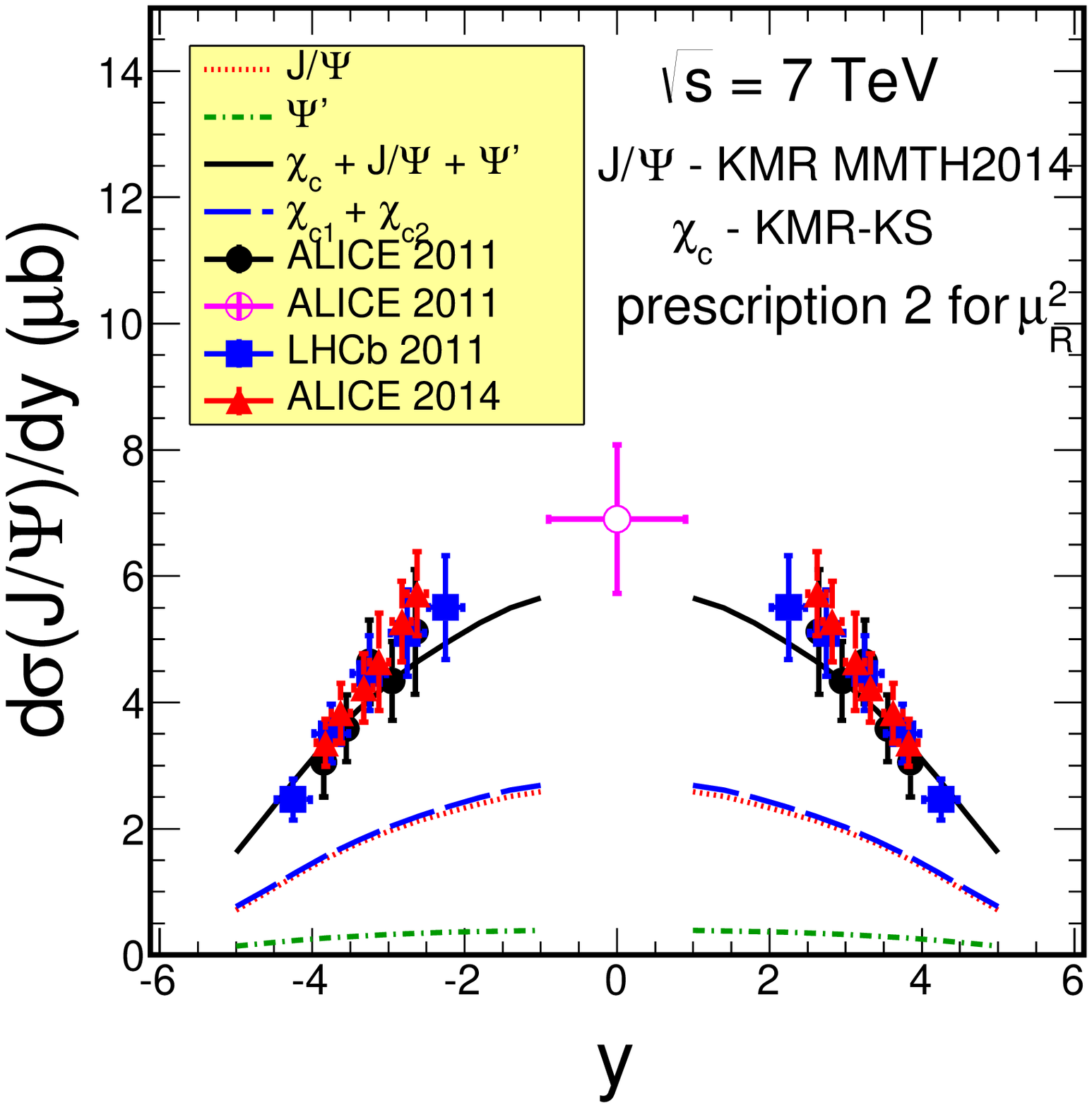}
\includegraphics[width=5.0cm]{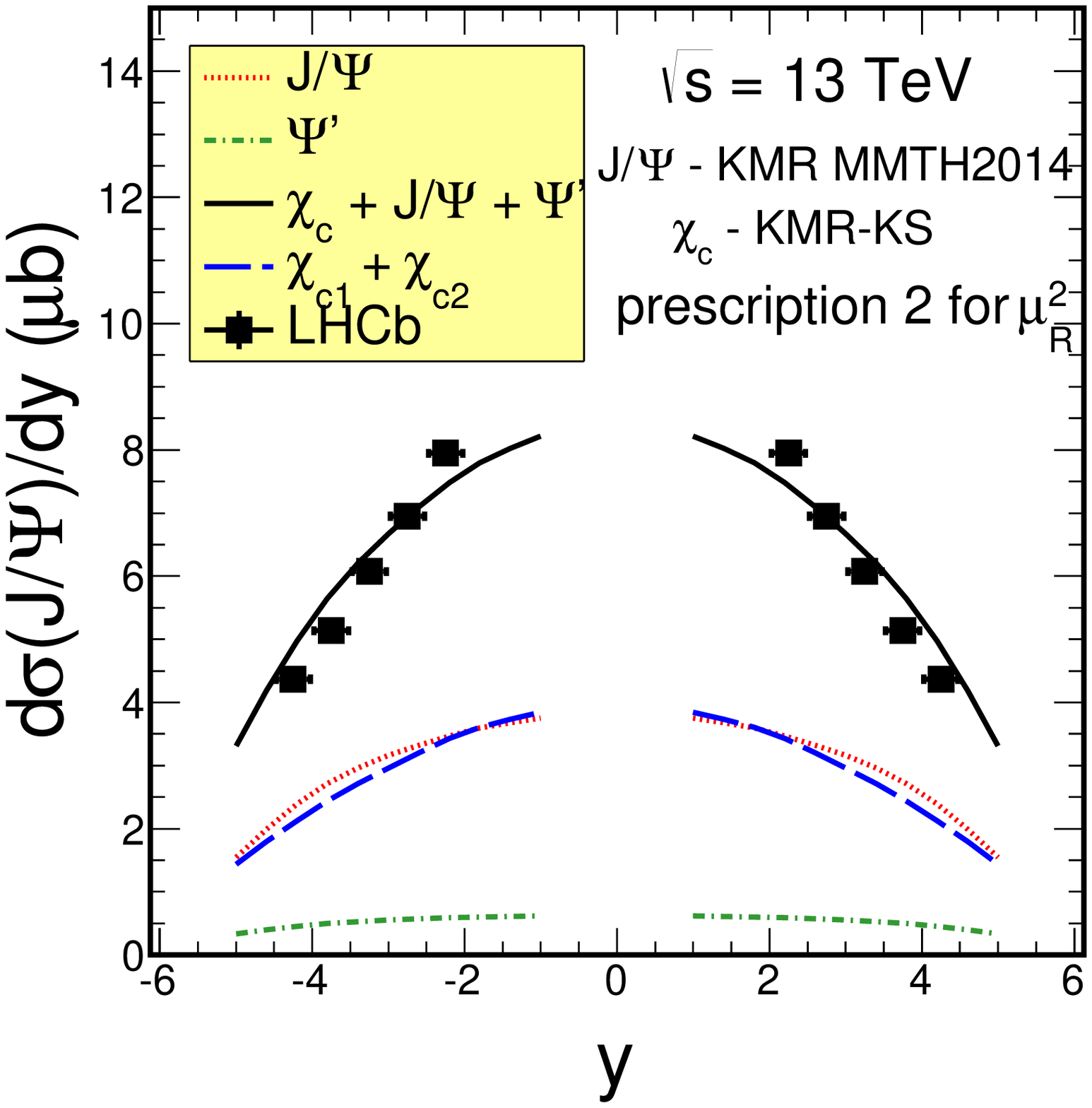}
\caption{
Rapidity distribution of $J/\psi$ mesons for all considered mechanisms
for three different energies. In this calculation only for
$\chi_{c}$ production the mixed UGDFs scenario was used.
}
\label{fig:dsig_dy_jpsi_all_mixed_KMR}
\end{figure}

Since, as discussed above, the longitudinal momentum fractions for $J/\psi$
and $\psi'$ are about order of magnitude larger than those
for $\chi_c$ production, we consider also a new scenario.
Here we take standard KMR UGDFs for the $S$-wave quarkonia and ``mixed''
UGDFs for the $\chi_c$ mesons. The resulting distributions are shown
in Fig.\ref{fig:dsig_dy_jpsi_all_mixed_KMR}. The agreement with the
experimental data is very good, but we cannot draw too 
strong conclusions. More systematic studies of low-$p_t$ distributions
of $J/\psi$ and $\chi_c$ mesons would certainly be very useful
in this context.

\section{Conclusions}

In the present paper we have focused on calculation of
cross sections for inclusive prompt production of $J/\psi$ and $\psi'$
in forward directions within the $k_t$-factorization approach.
In this calculation NR QCD matrix elements were used with
parameters of quarkonia $c \bar c$ wave functions at the origin 
taken from potential model(s).

In the present calculation we have used two different sets of
unintegrated gluon distribution functions: the Kimber-Martin-Ryskin UGDF
based on DGLAP collinear gluon distribution function, 
and the Kutak-Sta\'sto UGDF which includes nonlinear
effects at small $x$ values and describes exclusive production of 
$J/\psi$ \cite{CSS2015}.

We have included both direct component and the component related to
radiative decays of $\chi_c$ mesons. In general, they give similar
contributions for the integrated cross section.

We have compared our results with the recent results of
the ALICE and LHCb collaboration (small transverse momenta
and forward directions) at $\sqrt{s}$ = 7 TeV.
We have found that using standard KMR UGDF we overestimate the forward
production of $J/\psi$ in this case. The biggest contribution is given by
radiative decays of $\chi_{c}$ mesons. We have proposed how to modify UGDFs
to include possible onest of saturation effects.
In this mixed UGDF scenario, a reasonable description of the data is possible.
We have found that within model uncertainties (UGDFs, renormalization scale,
parameters of the nonrelativistic wave function) we can almost describe 
the production of $J/\psi$ or $\psi'$ at low transverse momenta and
forward direction including only color-singlet contribution.

We have discussed theoretical uncertainties related to the choice
of renormalization scales. In addition, we have discussed some open 
issues related to the KMR UGDFs. We have shown
how to modify the KMR UGDFs to include possible saturation effects.
A possible onset of saturation or in general nonlinear effects
for UGDFs was discussed, especially in the context of the LHCb data.
We have found that production of $\chi_{c}$ mesons in forward directions
is a very good way to search for the onset of saturation, because, as
discused in our paper, it probes smaller values of longitudinal
momentum fraction than the $J/\psi$ production and is therefore better
suited for that purpose.
Such an analysis could be done by the LHCb collaboration.

\vspace{1cm}


{\bf Acknowledgments}

We are indebted to Sergey Baranov, Roman Pasechnik, Vladimir Saleev 
and Wolfgang Sch\"afer for exchange of useful informations.
This work was partially supported by the Polish grant 
No. DEC-2014/15/B/ST2/02528 (OPUS)
as well as by the Centre for Innovation and Transfer of 
Natural Sciences and Engineering Knowledge in Rzesz\'ow.


\end{document}